\documentclass[journal]{IEEEtran}
\usepackage{amsmath,amsfonts}
\usepackage{algorithmic}
\usepackage{algorithm}
\usepackage{array}
\usepackage[caption=false,font=normalsize,labelfont=sf,textfont=sf]{subfig}
\usepackage{textcomp}
\usepackage{stfloats}
\usepackage{url}
\usepackage{verbatim}
\usepackage{graphicx}
\usepackage{cite}
\usepackage{booktabs}
\usepackage{multirow}
\usepackage{diagbox}
\usepackage{graphicx}
\usepackage{subfig}
\usepackage{amsfonts}
\usepackage{amsmath,amsfonts,amssymb,epsfig,cite,url,multirow,booktabs}
\usepackage{afterpage,mathrsfs,algorithm,algorithmic}
\usepackage{amsthm}
\usepackage{bbding}
\usepackage{xcolor}
\usepackage{colortbl}

\newtheorem*{remark}{\textbf{Remark}}

\usepackage[justification=centering]{caption}
\begin{document}

\title{Multi-Domain Supervised Contrastive Learning for UAV Radio-Frequency Open-Set Recognition}

\author{Ning Gao,~\IEEEmembership{Member,~IEEE,}~Tianrui Zeng, \IEEEmembership{Student Member,~IEEE,}~Bowen Chen, Donghong Cai, \IEEEmembership{Member,~IEEE,}\\Shi Jin,~\IEEEmembership{Fellow,~IEEE,} and~Michail Matthaiou,~\IEEEmembership{Fellow,~IEEE}

%\author{Ning Gao,~\IEEEmembership{Member,~IEEE,}~Le Liang,~\IEEEmembership{Member,~IEEE,}~Donghong Cai,~\IEEEmembership{Member,~IEEE,}\\ Xiao Li,~\IEEEmembership{Senior Member,~IEEE,}~and Shi Jin,~\IEEEmembership{Senior Member,~IEEE}
%\thanks{This work was supported in part by the National Science Foundation
%of China (NSFC) under Grants 62001109 \& 62261160576, and in part
%by the Start-up Research Fund of Southeast University under Grant 4009012307. (\emph{Corresponding author: Shi Jin.})}
\thanks{This work was supported in part by the National Key Research and Development Program of China under Grant 2024YFE0200700, in part by National Science Foundation of China (NSFC) under Grants 62371131, and in part by the program of Zhishan Young Scholar of Southeast University under Grant 2242024RCB0030.}
\thanks{N. Gao, T. Zeng and B. Chen are with the School of Cyber Science and Engineering, Southeast University, Nanjing 210096, China (e-mail: ninggao@seu.edu.cn; 220245586@seu.edu.cn; 213232141@seu.edu.cn).}

\thanks{D. Cai is with the College of Cyber Security, Jinan University, Guangzhou 510632, China (e-mail: dhcai@jnu.edu.cn).}
\thanks{S. Jin is with the National Mobile Communications
Research Laboratory, Southeast University, Nanjing 210096, China (e-mail: jinshi@seu.edu.cn).}% <-this % stops a space
\thanks{
M. Matthaiou is with the Centre for Wireless Innovation (CWI), Queen’s University Belfast, Belfast BT3 9DT, U.K. (e-mail: m.matthaiou@qub.ac.uk).}
}

\markboth{Manuscript}%
{Shell \MakeLowercase{\textit{et al.}}: A Sample Article Using IEEEtran.cls for IEEE Journals}


\maketitle

\begin{abstract}
5G-Advanced (5G-A) has enabled the vibrant development of low
altitude integrated sensing and communication (LA-ISAC) networks. As a core component of these networks, unmanned aerial vehicles (UAVs) have witnessed rapid proliferation in recent years. However, due to the lag in traditional industry regulatory norms, unauthorized flight incidents occur frequently, posing a severe security threat to LA-ISAC networks. To surveil the non-cooperative UAVs, in this paper, we propose a multi-domain supervised contrastive learning (MD-SupContrast) framework for UAV radio frequency (RF) open-set recognition. Specifically, first, the texture features and the time-frequency position features from the ResNet and the TransformerEncoder (TE) are fused, and then the supervised contrastive learning is applied to optimize the feature representation of the closed-set samples. Next, to surveil the invasive UAVs that appear in real life, we propose an improved generative OpenMax (IG-OpenMax) algorithm and construct an open-set recognition model, namely Open-RFNet. According to the unknown samples, we first freeze the feature extraction layers and then only retrain the classification layer, which achieves excellent recognition performance both in
closed-set and open-set recognitions. We analyze the computational complexity of the proposed model. Experiments are conducted with a large-scale UAV open dataset. The results show that the proposed Open-RFNet outperforms the existing benchmark methods in terms of recognition accuracy between the known and the unknown UAVs, as it achieves 95.12\% in closed-set and 96.08\% in open-set under 25 UAV types, respectively.
\end{abstract}

\section{Introduction}
\subsection{Background}
\IEEEPARstart{U}{nmanned} Aerial Vehicles (UAVs), which are unmanned aircrafts controlled by a wireless remote control device and their own program control device, have the advantages of small size, low cost and strong flexibility. Initially, UAVs were applied in military operations for reconnaissance and target strikes. With technological advancements and cost reduction, UAVs have gradually been applied in the civilian domain. Currently, the application scenarios of UAVs are becoming increasingly diverse, and they have been widely used in fields, such as forestry plant protection, communications and emergency rescue, etc~\cite{10004900,8873597,9762762}.

However, the contradiction between the advancement of the UAV technology and the regulatory perplexities has gradually exposed the security risks and management loopholes. Commercial UAVs are easy to purchase and develop, though most of them still do not have the online registration function. Furthermore, many of them do not have a geofencing or the geofencing option can be easily turned off by modifying their global positioning system (GPS). Since the flight rules of the UAVs are not easily followed by the malicious users, illegal invasion and privacy leakage become significantly likely in low altitude (LA) networks. In this case, the question of ``who is flying?" has to be urgently addressed. From an anti-UAV perspective, the UAV type recognition is a prerequisite technique to counter the invasion of the non-cooperative UAVs.

With the development of 5G-Advanced (5G-A) base stations, LA-integrated sensing and communication (ISAC) has become one of the promising technologies for the development of future wireless communications. The surveillance of LA UAVs is an important LA-ISAC application scenario. Traditionally, LA UAV surveillance mainly focuses on physical surveillance, which can be used for the surveillance of the flight trajectory for safety purposes. Therein, the physical states of the UAV are perceived via multi-modal sensing, including visual, acoustic and radar, etc. Nevertheless, the passive surveillance function for UAV types or identities, based on 5G-A base stations, has been less investigated.

\begin{figure}[!t]
	\centering
	\includegraphics[width=2.8in]{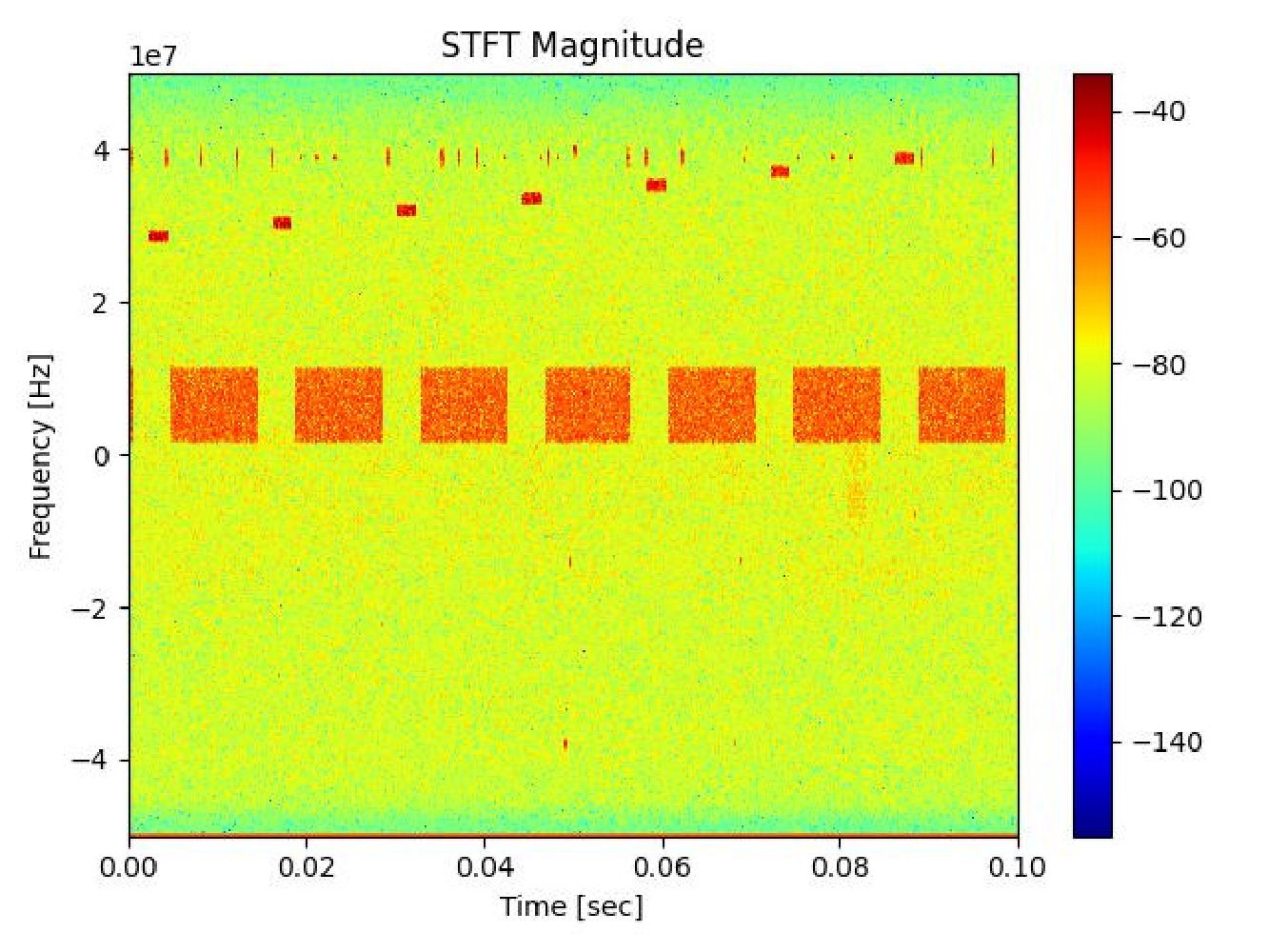}
	\caption{The schematic diagram of a UAV RF signal.}
	\label{fig_TFS}
\end{figure}

The radio frequency (RF) signal-based UAV recognition is a promising method to recognize UAV types, which is attributed to the following advantages. Compared with the radar and light detection, it is environmental friendly, since passive monitoring does not cause electromagnetic interference to the wireless environments and has low energy consumption. Compared with the vision and acoustic detection, it is environmentally robust, since it is insensitive to obstacles, non line-of-sight (NLoS) propagation conditions and the atmospheric sounds~\cite{rahman2024comprehensive}. The recognition performance is irrelevant to the size of the UAV, which makes it possible to identify even very covert illegal activities. The underlying principle of the RF signal-based UAV recognition is that different types of UAVs have different spectrum features in raw RF signals, such as the duty ratio, modulation pattern, which is regarded as an intrinsic signature of the UAV type. Figure~\ref{fig_TFS} shows a schematic diagram of the UAV RF signal of the popular DJI Phantom 4 Pro.
The large red rectangle represents the video transmission signal (VTS), which carries the UAV’s camera video stream.
The small red rectangle represents the control signal, which carries flight commands and logs.

\subsection{Motivation and Contribution}
There are some recent works that have studied the RF signal-based UAV recognition. The authors in \cite{al2019rf} proposed a deep neural network (DNN)-based method to detect the UAV presence, type and flight mode, where the results showed a general decline in performance when increasing the number of classes, i.e., the average accuracy decreased from 99.7\% for 2 classes to 46.8\% for 10 classes. The reason is that when the number of UAVs increase, the features among UAVs become extremely similar, whereas the signal of the commercial UAV is usually non-stationary and frequency hopping, which makes the basic CNNs struggle to classify. On the other hand, many works focus on the closed-set recognition, where the UAVs are categorized to the known classes under the assumption that the environment is completely closed. However, for illegal modified UAVs, the RF sample distribution is unknown, thus, there are significant limitations with closed-set recognition \cite{9517121,ARXIV25,tao2024threshold,yu2024open}. On the other hand, open-set techniques accept test samples from unknown classes, which requires the model to not only classify known classes but also detect unknown classes. This setting better reflects practical UAV surveillance environments, where many non-cooperative UAVs often do not appear in the training dataset. Recently, Tao \textit{et al}. proposed an open-set recognition method, based on a neural network and the OpenMax algorithm, for radar automatic UAV target recognition \cite{tao2024threshold}. Yu \textit{et al}. proposed a signal semantic-based open-set recognition which designs an outlier analysis-based semantic classifier to detect the unknown instances \cite{yu2024open}. However, the performance in open-set recognition is achieved at the expense of a performance reduction in closed-set recognition, where the recognition accuracies of some known classifications are significantly reduced. To the best of the authors' knowledge, the open-set recognition for non-cooperative UAVs has not yet been well studied.
These limitations in existing RF-based UAV recognition studies directly motivate our work.
Therefore, in this paper, we attempt to address the following problems: i) How to accurately recognize a UAV when the signal is subject to non-stationarities and frequency hopping? ii) How to improve the performance of open-set recognition while maintaining the performance of closed-set recognition as much as possible? To this end, we propose a multi-domain supervised contrastive learning (MD-SupContrast) framework to recognize the non-cooperative UAVs with RF signals. Specifically, both the texture feature and position feature are fused to represent the RF feature, while contrastive learning is used to optimize the feature representation. Then, the improved generative OpenMax (IG-OpenMax) algorithm is proposed to balance the recognition performance between the closed-set and the open-set. Our main contributions are summarized as follows:
\begin{itemize}
  \item An MD-SupContrast framework is proposed for UAV RF recognition, where the texture feature and the time-frequency position feature are jointly considered and fused via multiple non-linear layers (MNLs). Specifically, the texture feature is used to mitigate the negative impact from non-stationary signals and frequency hopping, while the position feature is used to enhance the recognition ability among known/unknown classes. In addition, to avail of texture features and time-frequency position features and avoid the unbalanced feature optimization, MD-SupContrast employs supervised contrastive learning to perform optimization with the feature representations of the samples themselves.
  \item Based on the proposed MD-SupContrast framework, we propose the Open-RFNet for recognizing the known and the unknown UAVs. In particular, to address the unbalanced recognition performance between the closed-set and the open-set, we propose a two-stage IG-OpenMax algorithm based on the generative OpenMax (G-OpenMax) algorithm. We first freeze the feature extraction layers of the trained closed-set model and then only retrain the classification layer for the unknown samples with a low training overhead.
  \item We adopt a large-scale UAV RF open dataset, namely DroneRFa\cite{dronerfa2024}, to evaluate the performance of the proposed Open-RFNet. The results show that the proposed Open-RFNet outperforms the benchmark methods in terms of recognition accuracy both in the known and the unknown UAVs.  The proposed Open-RFNet can achieve recognition accuracy of 95.12\% in closed-set and 96.08\% in open-set with 25 UAV types, respectively, while the performance gap between the closed-set and the open-set is only 0.96\%.
\end{itemize}
	
\subsection{Overview}
The rest of this paper is organized as follows: The related works are presented in Section \ref{sec:2}. The framework and the data preprocessing are presented in Section \ref{sec:3}. The closed-set recognition and the open-set recognition are proposed in Section \ref{sec:4} and Section \ref{sec:5}, respectively. The experimental results are showcased in Section \ref{sec:6}, while the paper is concluded in Section \ref{sec:7}.

\section{Related Work}
\label{sec:2}
\subsection{RF recognition methods}
RF recognition methods can accurately recognize the emitter target through RF signals. In the early stage, these methods relied on the manual feature extraction and expert knowledge. The high computational complexity makes it difficult to handle complex multi-dimensional features. For example, the expert knowledge would extract features by analyzing the deviation between the modulated signals and the ideal signals. These features include inphase/quadrature (I/Q) imbalance, amplitude error, and phase error, etc. Subsequently, machine learning methods, such as support vector machines (SVM) were typically used \cite{brik2008wireless}. However, this approach has some limitations as it relies on time-consuming feature engineering, thus being sensitive to environmental noise.

In recent years, deep learning has gradually been introduced into the RF recognition space \cite{9049161}. This effectively solves the generalization problem faced by traditional RF recognition methods in complex channel environments and provides strong support for large-scale device recognition tasks. For example, the authors in \cite{sankhe2019oracle} proposed the ORACLE method, which uses a convolutional neural network (CNN) to achieve high-precision classification. Reference \cite{peng2019deep} proposed the differential constellation trace figure (DCTF)-CNN method, which uses the DCTF to convert the differential relationship of signal time series into a two-dimensional image. Without carrier frequency and time synchronization, the CNN is directly employed to automatically learn the features, effectively reducing the recognition difficulty of hardware-similar ZigBee devices at low signal-to-noise ratio (SNR). With the popularity of large language models (LLMs), a LLM-enabled lightweight CNN framework was conceived in \cite{SCISGAOllm} to conduct the RF fingerprint identification for edge intelligence. In addition to the single modulation domain, feature extraction can be extended to multi-domain features, such as statistical, frequency domain, and wavelet domain \cite{9448105,lin2025,ezuma2019micro}.

RF recognition methods have been utilized in recognizing UAVs, but they are still relatively scarce. For example, a RF signal-based UAV surveillance system was developed in \cite{9768809}, with a high performance CNN which uses a structure of multi-level skip-connection and multi-level pooling. The authors in \cite{10513356} employed a lightweight backbone
network with multiscale convolution blocks to reduce the model size and enhance
the feature extraction ability, which can achieve an excellent accuracy at different SNRs.

\subsection{Other recognition methods}
UAV recognition could be also performed by leveraging vision, sound and radar-based methods. Vision-based recognition mainly relies on the image or video data stream captured by cameras. The authors in \cite{gokcce2015vision} used Haar-like features, local binary patterns (LBP), and histogram of oriented gradients (HOG) to extract different features from images, and constructs a cascaded classifier. However, the vision-based detection is sensitive to environment factors, including the lighting, weather and background noise, which degrade the recognition performance, especially of fast flying objects. Radar-based UAV recognition utilizes radio-echoes to detect the distant UAVs, where the position, speed, altitude, and the direction of the UAV can be determined. Compared with the vision-based methods, radar-based recognition is not limited by lighting conditions and can operate at night or in adverse weather. Reference \cite{de2018drone} was based on an X-band frequency modulated continuous wave (FMCW) radar. Through digital beamforming and 2D fast Fourier transform (FFT), it realized the detection of commercial UAVs within 2 km and has excellent range-speed association capability. However, radar-based UAV recognition has limited capabilities in detecting fine features, such as the UAV specific type. Moreover, the recognition performance for low-altitude, slow-moving and small-sized airborne targets is not very satisfactory. Sound-based UAV recognition captures sound signals generated during UAV flight, such as noise, rotor rotation sounds, and motor vibration sounds with microphone arrays. Then, it analyzes the acoustic features via signal processing and recognition algorithms to achieve UAV recognition and positioning. Reference \cite{al2019audio} proposed a method based on three neural networks, which realizes UAV detection and type recognition by extracting the acoustic fingerprint features and integrating data augmentation. However, this approach is highly susceptible to interference. Background noise in outdoor environments, such as wind, rain, traffic noise, biological sound and distance, may mask or distort the UAV's acoustic signals, which leads to a deterioration in the recognition accuracy.

\subsection{Open-set recognition}
Traditional machine learning and CNNs mainly focus on closed-set environments, which can only distinguish classifications that already exist in the training set. For the unknown classifications, the traditional methods mistakenly identify them as known classifications. Therefore, open-set recognition, which is closer to real-world scenarios, has been proposed \cite{scheirer2012toward,neira2018data,bendale2016towards,ge2017generative}. It requires the established model to not only distinguish classifications that appear in the training process but also effectively handle unseen classifications. Reference \cite{scheirer2012toward} proposed a 1-vs-set machine learning framework based on a linear kernel SVM, which restricts the decision boundaries of known classes to reduce the risk of open-set space by constructing a ``double-plane decision space". However, a single algorithm imposes certain limitations on the model. Then, \cite{neira2018data} adopted the idea of integration, fusing multiple algorithms to achieve open-set recognition. All the above methods are implemented based on traditional machine learning. For methods based on DNNs, the OpenMax was proposed in \cite{bendale2016towards}, which combines the Weibull distribution with the softmax layer to realize open-set recognition. This method successfully estimates the pseudo-probability of unknown classifications and achieves open-set recognition without considering the prior knowledge of unknown classifications. Reference \cite{ge2017generative} improved the OpenMax and then proposed the G-OpenMax, which synthesizes new sample classification via generative adversarial networks (GANs) to simulate real unknown category samples, thereby improving the recognition performance. However, this study has shown that although G-OpenMax can improve the performance of open-set recognition on monochromatic datasets, its performance improvement on natural images is not significant. Furthermore, there is still limited research on open-set recognition for UAVs \cite{tao2024threshold, yu2024open}.

\section{Preliminaries}
\label{sec:3}
In this section, we first propose the MD-SupContrast framework, and then formulate the signal preprocessing. Table~\ref{tab:notation} provides clear definitions of the mainly used symbols in this paper.
\begin{table}[t]
	\centering
	\caption{List of Symbols}
	\begin{tabular}{ll}
		\hline
		Symbol & Description \\
		\hline
		$s$ & Baseband I/Q transmitted signal \\
		$r$ & Received I/Q signal at the base station \\
		$d$ & Propagation distance \\
		$PL(d)$ & Distance-dependent path-loss function \\
		$\Delta d$ & Path perturbation caused by wobbling \\
		$\Delta d \sim \mathcal{N}(0,\sigma_d^2)$ & Gaussian model of wind-induced perturbation \\
		$\sigma_d^2$ & Variance of the perturbation process \\
		$\lambda$ & Signal wavelength \\
		$\eta$ & Additive white Gaussian noise (AWGN) \\
		$\mathcal{D}_\mathrm{train},\mathcal{D}_\mathrm{test}$ & Training and testing datasets \\
		$f(\cdot)$ & Feature extractor / encoder network \\
		$g(\cdot)$ & Classifier / decision function \\
		$z$ & Extracted fused deep feature representation \\
		$\tilde{z}_a$ & Texture-domain feature \\
		$\tilde{z}_b$ & Time-domain feature \\
		$\tilde{z}_c$ & Frequency-domain feature \\
		$\oplus$ & Concatenation operator for feature fusion \\
		$y, \hat{y}$ & Ground-truth label and predicted label \\
		$\tau$ & Temperature scaling factor for contrastive learning \\
		$\mathcal{L}_{sup}$ & Supervised contrastive loss \\
		$\mathcal{L}_{ce}$ & Cross-entropy loss \\
		$\mathcal{L}_{total}$ & Total loss integrating $\mathcal{L}_{sup}$ and $\mathcal{L}_{ce}$ \\
		$N$ & Number of known classes in open-set recognition \\
		$v_i(x)$ & Activation/logit vector of sample $x$ \\
		$\mu_j$ & Mean activation vector (MAV) of class $j$ \\
		$\rho_j$ & Weibull distribution parameters of class $j$ \\
		$\hat{P}(y=C \mid x)$ & Posterior probability of sample $x$ to class $C$ \\
		$\tilde{y}$ & Final predicted label (including the unknown class) \\
		\hline
	\end{tabular}
	\label{tab:notation}
\end{table}

\begin{figure*}[!t]
	\centering
	\includegraphics[width=0.86\textwidth]{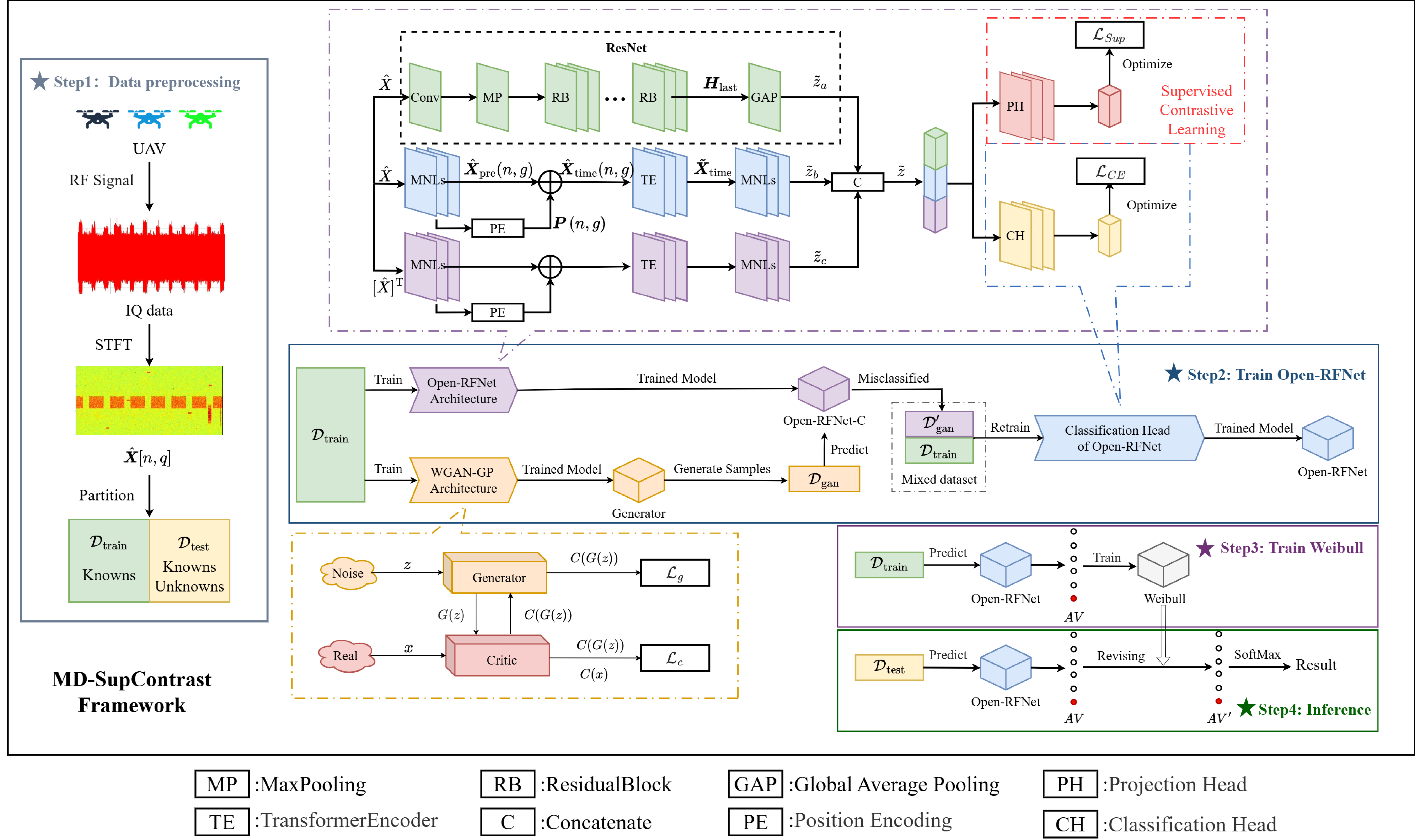}%
	\caption{The constructed Open-RFNet model based on the proposed MD-SupContrast framework.}
	\label{fig_model}
\end{figure*}
\subsection{Training Framework and Model}
We construct the Open-RFNet model based on the proposed MD-SupContrast framework, which is shown in Fig.~\ref{fig_model}. The proposed framework mainly consists of four stages: data preprocessing, Open-RFNet training, Weibull model training, and sample inference.
In the first stage, the collected UAV I/Q signals are sliced and denoised, and then transformed into time--frequency representations via the short-time Fourier transform (STFT). The obtained samples are subsequently partitioned into a training set $\mathcal{D}_{\mathrm{train}}$ and a test set $\mathcal{D}_{\mathrm{test}}$, which serve as the inputs for the subsequent stages.
In the second stage, the proposed Open-RFNet is trained on $\mathcal{D}_{\mathrm{train}}$ under supervised contrastive learning, and the resulting model is denoted as Open-RFNet-C, which is capable of performing closed-set recognition on the known classes. To further enable open-set recognition, we propose an IG-OpenMax algorithm. Specifically, a generator is trained based on the Wasserstein GAN with gradient penalty (WGAN-GP) framework to simulate unknown class samples, which is denoted as $\mathcal{D}_{\mathrm{gan}}$. The set $\mathcal{D}_{\mathrm{gan}}$ is fed into Open-RFNet-C for prediction, and the misclassified samples are collected to form $\mathcal{D}'_{\mathrm{gan}}$. The mixed dataset formed by $\mathcal{D}_{\mathrm{train}} \cup \mathcal{D}'_{\mathrm{gan}}$ is used to retrain the classification head while keeping the feature extractor frozen, yielding the final Open-RFNet model.
In the third stage, a Weibull distribution is fitted to the activation values produced by Open-RFNet on $\mathcal{D}_{\mathrm{train}}$ to perform statistical modeling. Finally, in the fourth stage, the trained Weibull calibration model is applied to revise the prediction results of samples in $\mathcal{D}_{\mathrm{test}}$, thereby enabling effective open-set recognition.

\subsection{Data Preprocessing}
\begin{figure}[!t]
	\centering
	\includegraphics[width=2.6in]{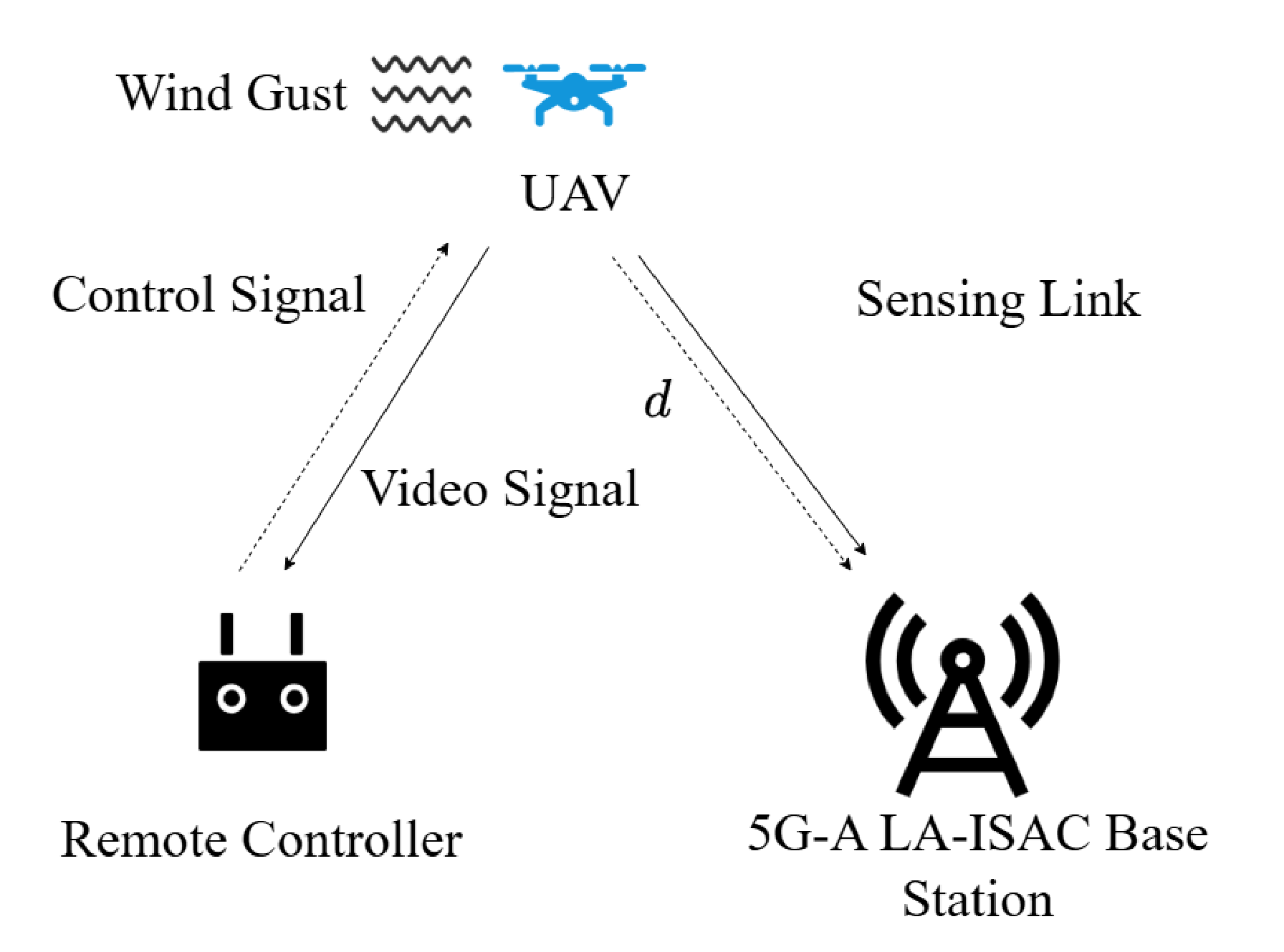}
	\caption{Air-to-ground LoS sensing link between the UAV and the base station.}
	\label{fig_system}
\end{figure}

The monitoring link between the UAV and the base station can be regarded as an air-to-ground channel, which is dominated by the line-of-sight (LoS) link, as illustrated in Fig.~\ref{fig_system}. On the other hand, various external factors, such as wind gust, may severely affect the phase of the propagation path, which is one of the important reasons making the signal non-stationary, especially in hover mode \cite{10857394}. Under these conditions, the I/Q signal transmitted over the wireless channel experiences attenuation and wobbling, and the demodulated I/Q signal received by the base station is
\begin{equation}r=\sqrt{PL(d)}\exp\left(-j\frac{2\pi}{\lambda}\big(d+\Delta d\big)\right)s+\eta,\end{equation}
where $d$ is the distance between the transmitter and the receiver, $PL(d)$ denotes the distance-dependent path-loss function between the UAV and the base station, which characterizes the large-scale fading over the air-to-ground link. The symbol $\Delta d$ is the phase affected by wobbling that is modeled as a zero-mean Gaussian random process, i.e., $\Delta d \sim \mathcal{N}(0, \sigma_d^2)$, where $\sigma_d^2$ denotes the variance of the random perturbation. The symbol $\lambda$ is the wavelength of the signal, $s$ denotes the baseband I/Q signal, and $\eta$ represents the additive white Gaussian noise (AWGN), which follows a zero-mean complex Gaussian distribution with variance $_{}\sigma^2$.

Since the UAV signals exhibit periodicity, there is a possibility of generating samples that contain only background noise without any UAV signals. Thus, when slicing the collected I/Q signals, we divide the original signal into sub-slices each with a size of 1/3 of the target slice. Subsequently, sub-slices with signal strength below a predefined noise threshold are filtered out. Finally, the retained sub-slices are concatenated sequentially to complete the denoising process. After slicing, the STFT is applied to each slice, where the STFT representation of the received I/Q signal is expressed as
\begin{equation}\boldsymbol{X}[n, q] = \sum_{m=0}^{M-1}r[m] \omega[n-m] \exp\left(-j 2\pi q m\right),\end{equation}
where $n$ is the time index, $q$ is the frequency index, $r[m]$ is the discrete time received signal, while $\omega[m]$ is the discrete window function with size $M$. The signal power spectrum of each time-frequency information can be given by
\begin{equation}\mathrm{dB}\big[\boldsymbol{X}[n, q]\big]=20\log_{10}\big|\boldsymbol{X}[n, q]\big|,\end{equation}
where $|\cdot|{}$ indicates the modulus of a complex number.
The placement of the UAVs at different locations makes the UAV signal power close to the RF receiver much larger than that of the remote UAVs. Thus, for the fairness of model training, we normalize each time-frequency information of the signal power spectrum to
\begin{equation}\hat{\boldsymbol{X}}[n, q]=\frac{\mathrm{dB}\big[\boldsymbol{X}[n, q] \big]-\min(\mathrm{dB}\big[\boldsymbol{X}(n,q)\big])}{\max(\mathrm{dB}\big[\boldsymbol{X}(n,q)\big])-\min(\mathrm{dB}\big[\boldsymbol{X}(n,q)\big])},\end{equation}
where $\min(\mathrm{dB}[\boldsymbol{X}(n,q)])$ and $\max(\mathrm{dB}[\boldsymbol{X}(n,q)])$ represent the minimum value and the maximum value of all elements in matrix $\mathrm{dB}[\boldsymbol{X}(n,q)]$, respectively.

\section{closed-set Recognition}
\label{sec:4}
In this section, we construct the feature extraction modules and the supervised contrastive learning, which is used for the UAV RF closed-set recognition. In general, supervised contrastive learning encourages samples from the same class to cluster together while pushing apart samples from different classes in the feature space, thereby enhancing the discriminability of RF representations \cite{khosla2020supervised}.
\subsection{Feature Extraction}
The ResNet-18 is used to extract texture features, which is primarily composed of 8 residual blocks with each block comprising two convolutional layers, a batch normalization layer, a ReLU activation function, and a shortcut connection. For the $l$-th residual block, the output can be expressed as
\begin{equation}
	\boldsymbol{H}_l = \mathrm{F}_l(\boldsymbol{I}_l;  \boldsymbol{W}_l) + \boldsymbol{I}_l,
\end{equation}
where $\mathrm{F}_l$ is the combination of the convolutional, the batch normalization, and the activation operations in the $l$-th residual block, $\boldsymbol{W}_l$ is the weight parameter of this block, and $\boldsymbol{I}_l$ is the input of this block.

Following the last residual block, its output $\boldsymbol{H}_\mathrm{last}$ is processed by the global average pooling (GAP) and a flatten layer to obtain the final texture feature representation, which is represented as
\begin{equation}	\tilde{\boldsymbol{z}}_\mathrm{a}=\mathrm{Flatten}\left(\text{GlobalAvgPooL}\left(\boldsymbol{H}_\mathrm{last}\right)\right),
\end{equation}
where $\text{GlobalAvgPooL}(\cdot){:}~\mathbb{R}^{c\times d_1\times d_2}\to\mathbb{R}^{c\times 1}$ is the global average pooling, while symbol $\mathrm{Flatten}(\cdot){:}~\mathbb{R}^{c\times 1}\to\mathbb{R}^{c}$ represents the flattening operation. Here, $c$ denotes the number of feature channels, while $d_{1}$ and $d_{2}$ represent the height and width
of the feature map, respectively. The TE is used to extract the time-frequency position features. Therein, the time domain position features are extracted by using the time-frequency signal $\hat{\boldsymbol{X}}(n,q)$, while the frequency domain position features are extracted by using the transpose of $\hat{\boldsymbol{X}}(n,q)$, i.e., $\hat{\boldsymbol{X}}^\mathrm{T}(n,q)$.

To extract the time domain position features, MNLs are first employed to perform preliminary feature extraction on $\hat{\boldsymbol{X}}(n,q)$, yielding $\hat{\boldsymbol{X}}_\mathrm{pre}(n,g)$. Subsequently, the position encoding is incorporated into the preliminary features to distill the time domain position features, which are denoted as
\begin{equation}
	\hat{\boldsymbol{X}}_{\mathrm{time}}(n,g)=\hat{\boldsymbol{X}}_\mathrm{pre}(n,g)+\boldsymbol{P}\left(n,g\right),
\end{equation}
\begin{equation}\boldsymbol{P}\left(n,g\right)=
	\begin{cases}
		\sin(\theta_gn), \mathrm{if}\ g\%2=0, \\
		\cos(\theta_gn), \mathrm{if}\ g\%2=1,
	\end{cases}
\end{equation}
where $g$ is the feature dimension index, while the total number of feature dimensions is $\tilde{G}$. The symbol $\boldsymbol{P}(\cdot)$ is the position encoding operation, $\hat{\boldsymbol{X}}_{\mathrm{time}}(n,g)$ is the preliminary position features in time domain, $\%$ represents the remainder operation, and $\theta_g$ is the hand crafted frequency, i.e., $\theta_g=10000^{-\frac{2g}{\tilde{G}}}$.

TE is based on the self-attention mechanism, which allows the model to flexibly focus on different positions in the input sequence, enabling the capture of long and short distance dependencies and performs well in learning long range dependencies and interactions in sequence data. In this mechanism, each word in the input sequence calculates its correlation with other words in the self-attention layer through the query ($\boldsymbol{Q}$), key ($\boldsymbol{K}$), and value ($\boldsymbol{V}$) vectors, thereby generating a weighted output. Note that $\boldsymbol{Q}$ represents the information that the current word needs to focus on, while the $\boldsymbol{K}$ vector represents the features of each word, which is used to compare with the $\boldsymbol{Q}$ vectors of other words to determine their correlations, and the value vector represents the actual content of each word. The weights of all words are assigned according to the similarity between $\boldsymbol{Q}$ and $\boldsymbol{K}$ to generate the output. The TE module is composed of multiple stacked TransformerEncoders. The $l$-th TE can be given by \cite{yu2024open}
\begin{equation}
	\varphi_l(\boldsymbol{Y}_l)=\mathrm{softmax}\left(\frac{\boldsymbol{Y}_l\boldsymbol{Q}_l\boldsymbol{K}_l^\mathrm{T}\boldsymbol{Y}_l^\mathrm{T}}{\sqrt{\gamma_l}}\right)\boldsymbol{Y}_l\boldsymbol{V}_l,
\end{equation}
where $\boldsymbol{Y}_l$ denotes the input to the $l$-th TE. For the first TE, i.e., $l=0$, $\boldsymbol{Y}_0$ is equivalent to $\hat{\boldsymbol{X}}_{\mathrm{time}}(n,g)$, $\mathrm{softmax}(\cdot)$ represents the softmax activation function, $\boldsymbol{Q}_l,\boldsymbol{K}_l,\boldsymbol{V}_l\in\mathbb{R}^{n\times \gamma}$ are learnable parameters, and $\gamma$ is the size of the mapping dimension.
Therefore, the output of the TE module can be given by
\begin{equation}\tilde{\boldsymbol{X}}_{\mathrm{time}}=\varphi_{L-1}\left(\varphi_{L-2}\big(\ldots(\varphi_{1}(\varphi_{0}(\hat{\boldsymbol{X}}_{\mathrm{time}}(n,g)))\big)\right),\end{equation}
where $L$ is the number of TransformerEncoders. After that, the output is flattened and further mapped by MNLs to obtain the time domain position features, which are expressed as
\begin{equation} \tilde{\boldsymbol{z}}_\mathrm{b}=\mathrm{MNLs}\left(\mathrm{Flatten}\big(\tilde{\boldsymbol{X}}_{\mathrm{time}}\big)\right).
\end{equation}

On the other hand, the same procedure is applied to the transpose of $\hat{\boldsymbol{X}}(n,q)$ to obtain the frequency domain position features $\tilde{\boldsymbol{z}}_\mathrm{c}$.

After extracting the texture features and the time-frequency position features, we use MNLs to fuse the multi-domain features to obtain the fused feature representation, as follows
\begin{equation}\tilde{\boldsymbol{z}}=\mathrm{MNLs}\big((\tilde{\boldsymbol{z}}_\mathrm{a}\oplus\tilde{\boldsymbol{z}}_\mathrm{b}\oplus\tilde{\boldsymbol{z}}_\mathrm{c},)\big),\end{equation}
where  $\oplus$ represents the concatenation operation, $\tilde{\boldsymbol{z}}_\mathrm{a}$ denotes the texture features, $\tilde{\boldsymbol{z}}_\mathrm{b}$ and $\tilde{\boldsymbol{z}}_\mathrm{c}$ denote the time-domain position features and frequency-domain position features, respectively.
\begin{remark}
As shown in Fig.~\ref{fig_TFS}, the texture features focus on the basic geometric features of the red matrix, such as the length-to-width ratio of the rectangle, the sharpness of the edge contour, and the spatial distribution pattern of the internal amplitude, including the uniformity and gradient trend of the energy amplitude. The texture features are local features which have no global statistical characteristics, thereby reducing the negative impact of non-stationary UAV signals and frequency hopping in VTS. On the other hand, the model uses the TE module to extract position features in the time-frequency domain. This emphasizes the specific location of the red matrix in these domains and its relative position to other signal blocks globally. These are global features which have the profile features of UAV types, thereby enhancing the UAV RF recognition performance.
\end{remark}
\subsection{Supervised Contrastive Learning}
The supervised contrastive learning is used to optimize the fused features, enabling effective utilization between features from different domains. Compared with unsupervised contrastive learning, supervised contrastive learning makes full use of label information. It can use multiple samples of the same class as positive sample pairs, rather than relying only on the data augmentation to generate positive samples, which helps to better learn the consistency within the classification. In addition, supervised contrastive learning employs the SupCon loss function, which overcomes various shortcomings of the cross-entropy loss function, such as lack of robustness to noisy labels, insufficient margin between the classes, and unbalanced optimization of multiple features \cite{chen2020simple,bi2022vision,liu2016large,peng2022balanced}. These advantages contribute to recognizing the type of UAVs from the extremely similar and noisy I/Q signals.

Specifically, given a batch of input samples, random data augmentation is performed to generate two different signal views. These two views pass through the previously mentioned feature extraction network to obtain two feature representations. Finally, the supervised contrastive loss function is calculated based on these feature representations. The main idea is to regard all samples of the same category as positive samples and reduce the distance between them, while regarding samples of other classifications as negative samples and increasing the distance between them.
In supervised contrastive learning, each sample in the batch is sequentially taken as an anchor sample, and its corresponding positive and negative sample sets are constructed according to the class labels.
The supervised contrastive loss function is denoted as
\begin{equation}\mathcal{L}_{\mathrm{sup}}
	=\sum_{i\in \boldsymbol{I}}-\frac{1}{|\boldsymbol{P(i)}|}\sum_{p\in \boldsymbol{P(i)}}\log\frac{\exp\left(\boldsymbol{z}_i\cdot\frac{\boldsymbol{z}_p}{\mathcal{T}}\right)}{\sum_{a\in \boldsymbol{A(i)}}\exp\left(\boldsymbol{z}_i\cdot\frac{\boldsymbol{z}_a}{\mathcal{T}}\right)},\end{equation}
where $\boldsymbol{I}$ is the set of all anchor samples, and $i$ is a single anchor sample; $\boldsymbol{P(i)}$ is the set of all positive samples of the anchor sample $i$,  and $\boldsymbol{A(i)}$ is the set of all negative samples of the anchor sample $i$. Moreover, $\boldsymbol{z}_{i}$, $\boldsymbol{z}_{p}$, $\boldsymbol{z}_{a}$ are the projection vectors corresponding to the anchor sample, positive samples, and negative samples, respectively, while $\mathcal{T}$ represents the temperature parameter.
\begin{remark}
The Open-RFNet model fuses texture features and time-frequency position features to obtain the fused feature representation. The traditional cross-entropy loss function is prone to the unbalanced optimization problem for multiple features. This is because it only focuses on the matching degree between the overall prediction result and the real label. When a certain feature, such as texture feature, has better prediction performance, it contributes more to the final prediction, and the corresponding gradient dominates in backpropagation. Then, the model can prioritize optimizing the parameters of this dominant feature. Nevertheless, supervised contrastive learning is used to optimize feature representations at the feature level. This not only enhances intra-class consistency but also expands inter-class differences, and avoids such unbalanced feature optimization problems.
\end{remark}
\section{Open-set Recognition}
\label{sec:5}
In this section, we first propose the two-stage IG-OpenMax algorithm and then discuss the training details of the GAN for the UAV RF open-set recognition.

\subsection{Problem Formulation of Open-set Recognition}

Let $\mathcal{Y}_K=\{1,2,\dots,N\}$ denote the set of known classes (KKCs) accessible during the training phase, and let $\mathcal{Y}_U$ denote the set of unknown classes (UUCs) that do not appear in training but may occur during testing.
Based on these label spaces, we define the corresponding sample domains as
\begin{equation}
	\mathcal{D}_K = \{(x,y)\mid y\in\mathcal{Y}_K\},
\end{equation}
\begin{equation}
	\mathcal{D}_U = \{(x,y)\mid y\in\mathcal{Y}_U\}.
\end{equation}

Accordingly, the training and testing datasets satisfy the following inclusion relations:
\begin{equation}
	\mathcal{D}_{\mathrm{train}} \subseteq \mathcal{D}_K,
\end{equation}
\begin{equation}
	\mathcal{D}_{\mathrm{test}} \subseteq \mathcal{D}_K \cup \mathcal{D}_U.
\end{equation}

The goal of open-set recognition is to learn a classifier $f$ that simultaneously achieves correct identification of known class samples and reliable rejection of unknown class samples. Formally, $f$ should satisfy
\begin{equation}
	\forall (x,y)\in \mathcal{D}_{\mathrm{test}} \cap \mathcal{D}_K:\quad f(x)=y,
\end{equation}
\begin{equation}
	\forall (x,y)\in \mathcal{D}_{\mathrm{test}} \cap \mathcal{D}_U:\quad f(x)=\mathrm{unknown}.
\end{equation}

In other words, the classifier is required to output the correct class label for test samples belonging to $\mathcal{D}_K$ and assign all samples originating from $\mathcal{D}_U$ to an additional unknown class.

\subsection{IG-OpenMax}
Open-RFNet implements the IG-OpenMax algorithm to detect illegal UAVs, which is an improved version of the G-OpenMax algorithm. Similar to G-OpenMax, the IG-OpenMax algorithm also introduces unknown class sample data during closed-set recognition training by providing explicit probability estimation for the unknown classes. The explicit representation of unknown classes enables the classifier to leverage the knowledge of both known and unknown samples to locate decision boundaries more effectively, thereby achieving clearer differentiation between known and unknown classes. Specifically, a conditional deep convolutional generative adversarial network (cDCGAN) is first employed to synthesize labeled samples for each known class.
Given a random noise vector $z$ and the encoded label information $e_g(y)$ of a known class $y \in \mathcal{Y}_K$, the generator produces
\begin{equation}
	\tilde{x} = G\big([z, e_g(y)]\big),
	\qquad
	\tilde{y} = y,
\end{equation}
where $G$ is the generator model, $z$ is sampled from a prior distribution, and $e_g(y)$ denotes the label embedding used for conditioning.
All such generated labeled samples form the synthetic dataset
\begin{equation}
	\mathcal{D}_{\mathrm{gan}}
	=
	\big\{
	(\tilde{x}, \tilde{y})
	\,\big|\,
	\tilde{x} = G([z, e_g(y)]),\;
	y \in \mathcal{Y}_K
	\big\}.
\end{equation}

These generated samples are then fed into the baseline model Open-RFNet-C with classifier $f_C(\cdot)$ for closed-set evaluation.
The subset of generated samples that are misclassified by Open-RFNet-C is collected as simulated unknown samples:
\begin{equation}
	\mathcal{D}_{\mathrm{gan}}'
	=
	\big\{
	(\tilde{x}, \mathrm{unknown})
	\,\big|\,
	(\tilde{x}, \tilde{y}) \in \mathcal{D}_{\mathrm{gan}},
	\;
	f_C(\tilde{x}) \neq \tilde{y}
	\big\}.
\end{equation}

These misclassified samples serve as approximations to real unknown class data and are used to expand the open-set training set:
\begin{equation}
	\mathcal{D}_{\mathrm{train}}^{\mathrm{OSR}}
	=
	\mathcal{D}_{\mathrm{train}}
	\cup
	\mathcal{D}_{\mathrm{gan}}'.
\end{equation}

Unlike the traditional G-OpenMax algorithm, the proposed IG-OpenMax freezes the feature extraction part of Open-RFNet-C and only retrains the classification head.
This design is adopted because retraining the entire network would destroy the original feature distribution, causing the generated samples to deviate from real unknown class samples in the new feature space.

After training the classier, the prediction results of the known and unknown class samples from Open-RFNet are fed into the OpenMax layer to obtain the open-set recognition result. Specifically, for each known category $j$, we first calculate the activation vectors (AV) of all training samples that are correctly recognized as this classification, as follows
\begin{equation}  \label{eq:7}
	\mathcal{S}_j = \left\{ \boldsymbol{v}_i(x_i) \mid \hat{y}_i = y_i = j, \, x_i, y_i \in \mathcal{D}_\mathrm{train} \right\},
\end{equation}
where $\boldsymbol{v}_i(x_i)=\left( v_{i,0}(x),\ v_{i,1}(x), \dots,\ v_{i,N}(x) \right)$ consists of the logits from this sample to all classifications with the number of the known classifications $N$, while $\mathcal{D}_\mathrm{train}$ represents the training set. Denoting the mean activation vector (MAV) of the $j$-th class as $\boldsymbol{\mu}_{j}$, then we use the Weibull distribution to fit the distance of each classification, according to
\begin{equation}  \label{eq:6}
	\rho_j=\begin{pmatrix}\tau_j,\kappa_j,\chi_j
	\end{pmatrix}
	=\mathrm{FitHigh}\!\left(\|{\mathcal{S}_j-\boldsymbol{\mu}_{j}} \|,\, L_{\text{tail}}\right),
\end{equation}
where $\tau_j$ is the shape parameter, $\kappa_j$ is the scale parameter,
$\chi_j$ is the location parameter and $L_{\text{tail}}$ represents the tail length hyperparameter controlling the number of extreme samples used for fitting. Moreover, $\mathrm{FitHigh}(\cdot)$ denotes the Weibull tail-fitting function, commonly used in OpenMax-based open-set recognition, which fits a Weibull distribution to the largest $L_{\text{tail}}$ activation distances of class $j$ to model its tail behavior. For a test sample, we first compute $\boldsymbol{v}_i(x)$ and then adjust the AVs with
\begin{equation} \label{eq:5}
	\tilde{\boldsymbol{v}}_i(x)=\boldsymbol{v}_i(x)\circ\boldsymbol{c}_j(x)=\left( \tilde{v}_{i,0}(x),\ \tilde{v}_{i,1}(x), \dots,\ \tilde{v}_{i,N}(x) \right),
\end{equation}
where ${\boldsymbol{v}_i(x)\circ\boldsymbol{c}_j(x)}$ represents the element-wise (Hadamard) multiplication between the original activation vector and the calibration coefficient vector. The term $\left( \tilde{v}_{i,0}(x),\, \tilde{v}_{i,1}(x),\, \dots,\, \tilde{v}_{i,N}(x) \right)$ is the explicit expansion of the adjusted activation vector, which lists all of its $(N+1)$ components.
The decrements sum of the original scores is used to derive the score of the unknown classification, which is given by

\begin{equation} \label{eq:1}
	\tilde{v}_{i,N+1}(x) = \sum_{j=0}^{N} \big({v}_{i,j}(x) - \tilde{{v}}_{i,j}(x) \big),
\end{equation}
where $\boldsymbol{c}_j(x)$ is the correction weight vector generated for samples in class $j$ based on the fitting results of the Weibull distribution, which only corrects the top-$\alpha$ classifications with the highest probabilities. For the AVs of the known classes, we have
\begin{equation} \label{eq:3}
	\boldsymbol{c}_{\boldsymbol{s}(k)}(x) = 1 - \frac{\alpha-k}{\alpha}\exp\left(-\left(\frac{\|x-\tau_{s(k)}\|}{\chi_{s(k)}}\right)^{\kappa_{s(k)}}\right),
\end{equation}
whereas for the unknown class, we have
\begin{equation}  \label{eq:4}
	\boldsymbol{c}_{\boldsymbol{s}(k)}(x)=1-\frac{\alpha-k}{\alpha},
\end{equation}
where $\boldsymbol{s}(k)=\mathrm{argsort}(\boldsymbol{v}_i(x))$ denotes the index corresponding to the
$k$-th largest probability classification.
Finally, we perform a softmax operation on the scores of all classifications and the score of the unknown classification to obtain the final open-set recognition distribution, which is given by
\begin{equation} \label{eq:2}
	\hat{P}(y=\mathcal{C}|x)=\frac{\exp\big(\tilde{v}_{i,\mathcal{C}}(x)\big)}{\sum_{j=0}^{N+1}\exp\big(\tilde{v}_{i,j}(x)\big)}.
\end{equation}

Unlike traditional threshold-based open-set methods, OpenMax does not rely on a manually fixed confidence threshold for rejection. Instead, it estimates the unknown probability through Weibull fitting based on extreme value theory (EVT). In our implementation, the only tunable parameter is the tail size~$L_{\text{tail}}$, which controls the number of extreme distances used for Weibull fitting of each class. This parameter determines how much probability mass is redistributed from the known classes to the unknown class, thus affecting the trade-off between known class accuracy and unknown class accuracy. We validate multiple $L_{\text{tail}}$ values on the validation set and select the one that yields a balanced performance between closed-set and open-set performances.
\begin{remark}
In the traditional G-OpenMax framework, the entire model, which includes the feature extractor, is retrained after incorporating the generated samples. This retraining substantially alters the original feature space learned by the closed-set model Open-RFNet-C. In the newly formed feature space, the distributional discrepancy between the generated samples and real data becomes more pronounced, enabling the model to easily distinguish between them. As a result, the generated samples tend to be separated from real samples, including real unknown class samples, rendering them ineffective for approximating the unknown class distribution.
In contrast, the proposed IG-OpenMax freezes the feature extractor during the second training stage and updates only the classification head, thus preserving the original feature space. Within this unchanged feature space, the misclassified generated samples typically lie in boundary regions, where real unknown class samples are also likely to appear. Consequently, these generated samples can more accurately approximate the distribution of real unknown classes in the preserved feature space, thereby substantially improving the open-set recognition performance.
\end{remark}
\begin{algorithm}
	\caption{IG-OpenMax for Open-Set Recognition}
	\begin{algorithmic}[1]
		\STATE \textbf{Input:} Training dataset $\mathcal{D}_\mathrm{train}$ consisting of known class samples,
		test dataset $\mathcal{D}_\mathrm{test}$ consisting of both known class and unknown class samples,
		number of known classifications $N$,
		number of top-score predictions to calibrate $\alpha$.
		\STATE \textbf{Stage-I: Train}
		\STATE Initialize Open-RFNet as Open-RFNet-B with feature extraction module $\mathcal{F}$, classification head $\mathcal{C}$, and projection head $\mathcal{P}$, that is $\text{Open-RFNet-B} = \{\mathcal{F},\mathcal{C}, \mathcal{P}\}$;
		\STATE Train Open-RFNet-B from $\mathcal{D}_\mathrm{train}$;
		
		\STATE Train a  DCGAN model $G$ from $\mathcal{D}_\mathrm{train}$;
				
		\STATE Select simulated unknown samples $\mathcal{D}_{\mathrm{simu}}$ with Open-RFNet-B;

		\STATE Construct Open-RFNet by freezing $\mathcal{F}$ and retraining $\mathcal{C}$ on $\mathcal{D}_\mathrm{train} \cup \mathcal{D}_{\mathrm{simu}}$;
		
		\FOR{each sample $x_i \in \mathcal{D}_\mathrm{train}$}
		\STATE $\boldsymbol{v}_i(x), \hat{y}_i = \text{Open-RFNet}(x_i)$;
		\ENDFOR
		\FOR{each known classification $j \in \{1,\dots,N\}$}

		\STATE Calculate $\mathcal{S}_j$ with Eq.~(\ref{eq:7});
		\STATE Calculate MAV, $\boldsymbol{\mu}_{j} = \mathrm{mean}(\mathcal{S}_j)$;
		\STATE Calculate $\rho_j$ with Eq.~(\ref{eq:6});
		\ENDFOR
		
		\STATE \textbf{Stage-II: Test}
		\FOR{each sample $x_i \in \mathcal{D}_\mathrm{test}$}
		\STATE $\begin{aligned}
			\boldsymbol{v}_i(x), \hat{y}_i = \text{Open-RFNet}(x_i);
		\end{aligned}$
		\STATE Let $\boldsymbol{s}(k)=\mathrm{argsort}(\boldsymbol{v}_i(x))$,
		Let $\boldsymbol{c}_j(x) = 1$;
			\FOR{$k = 1, \dots, \alpha$}
				\IF {$\boldsymbol{s}(k) \neq N+1$}
					\STATE Calculate $\boldsymbol{c}_{\boldsymbol{s}(k)}(x)$ with Eq.~(\ref{eq:3});
					
				\ELSE
					\STATE Calculate $\boldsymbol{c}_{\boldsymbol{s}(k)}(x)$ with Eq.~(\ref{eq:4});
				\ENDIF
			\ENDFOR
		\STATE Calculate $\tilde{\boldsymbol{v}}_i(x)$ with Eq.~(\ref{eq:5});
		\STATE Calculate $\tilde{v}_{i,N+1}(x)$ with Eq.~(\ref{eq:1})
		and calculate $\hat{P}(y=\mathcal{C}|x)$ with Eq.~(\ref{eq:2});
		\STATE Let $\tilde{y} = \mathrm{argmax}(\hat{P}(y=\mathcal{C}|x))$;
		\STATE \textbf{Output:} Open-set recognition result $\tilde{y}$.
		\ENDFOR
	\end{algorithmic}
	\label{algorithmic_1}
\end{algorithm}
\subsection{GAN Training}
We use the WGAN-GP framework to train a generator based on the cDCGAN architecture. By introducing a gradient penalty, WGAN-GP effectively alleviates the training instability of the original WGAN \cite{gulrajani2017improved}.
The WGAN utilizes weight clipping to ensure the Lipschitz constraint of the discriminator,
which is referred to as the critic.
Different from the WGAN, the WGAN-GP enforces the Lipschitz constraint by penalizing the gradient norm of the critic with respect to the input.
Specifically, the objective function of the WGAN-GP adds a gradient penalty term on the basis of the original WGAN,
and the critic loss can be expressed as

\begin{equation}
	\begin{aligned}
		\mathcal{L}_{c}
		&= \mathbb{E}_{\tilde{x}, \tilde{y} \sim \mathbb{P}_{g}}[C(\tilde{x}, \tilde{y})]
		- \mathbb{E}_{x, y \sim \mathbb{P}_{r}}[C(x, y)] \\
		&\quad + \beta \mathbb{E}_{\hat{x}, \hat{y} \sim \mathbb{P}_{\hat{x}}}
		\left[
		\left(\|\nabla_{\hat{x}} C(\hat{x}, \hat{y})\|_{2} - 1\right)^{2}
		\right].
	\end{aligned}
\end{equation}

For the generator, its objective is to maximize the critic's estimation for the generated samples,
which corresponds to minimizing the following generator loss:

\begin{equation}
	\mathcal{L}_{g} =
	- \mathbb{E}_{\tilde{x}, \tilde{y} \sim \mathbb{P}_{g}}[C(\tilde{x}, \tilde{y})],
\end{equation}
where $\mathbb{E}_{x \sim \mathbb{P}}[\cdot]$ denotes the expectation taken over the random variable $x$ drawn from the probability distribution $\mathbb{P}$.
$\mathbb{E}_{\tilde{x}, \tilde{y} \sim \mathbb{P}_{g}}[C(\tilde{x}, \tilde{y})]
- \mathbb{E}_{x, y \sim \mathbb{P}_{r}}[C(x, y)]$ is the objective function of the original WGAN, which represents the difference between the expected values of the generated samples and the real samples in the critic,  respectively. The expression $ \beta\mathbb{E}_{\hat{x}, \hat{y} \sim \mathbb{P}_{\hat{x}}}\left[\left(\|\nabla_{\hat{x}}C(\hat{x}, \hat{y})\|_{2} - 1\right)^{2}\right]$ is the gradient penalty term, which is used to ensure that the gradient norm of the critic is close to 1, thus satisfying the Lipschitz constraint. Therein, $C$ is the critic model, $\beta$ is the coefficient of the gradient penalty, $x$, $y$ are the real samples and their corresponding labels, whereas $\tilde{x}$, $\tilde{y}$ are the generated samples and their corresponding labels. Specifically,
$\tilde{x}=G\big(\left[z,e_{g}(y)\right]\big)$ and $\tilde{y}=y$, where $G$ is the generator model, $z$ is the input random noise, and $e_{g}(y)$ is the result of the generator encoding the label information. The symbols $\hat{x}$, $\hat{y}$ are the random interpolation points between the real samples and the generated samples, such that
$\hat{x}=\epsilon x+(1-\epsilon)\tilde{x}$
and
$\hat{y}=y,$
where $\epsilon$ is a random number uniformly distributed in the interval $[0,1]$.  We summarize the proposed IG-OpenMax algorithm in \textbf{Algorithm \ref{algorithmic_1}}.

\section{Complexity Analysis}
\label{sec:6}
Firstly, we analyze the time complexity of the Open-RFNet for closed-set recognition, which can be divided into two parts: the texture feature extraction part and the positional feature extraction part. Texture features are extracted through a ResNet network, wherein the time complexity mainly comes from the convolutional layers, batch normalization layers, and pooling layers. In the ResNet-based texture feature extractor, each convolutional unit consists of a convolution layer followed by a batch normalization layer. We denote by $U$ the total number of such convolution--batch normalization units, while the index $u = 1,\ldots,U$ is used to enumerate them in the subsequent complexity analysis. We assume that the batch size is $B$, the number of input channels is $C_\mathrm{in}$, the number of output channels is $C_\mathrm{out}$, the size of the input feature map is $H\times W$. We assume that all convolution kernels and pooling windows are square, where $K_\mathrm{conv}$, $K_\mathrm{avg}$ and $K_\mathrm{max}$ denote the side lengths of the convolution kernel, average pooling window, and max pooling window, respectively.
Then, the time complexity of the convolutional layers can be represented as
$\mathcal{O}(C_\mathrm{in}\times C_\mathrm{out}\times K_\mathrm{conv}^2\times H\times W)$,
the batch normalization layers can be represented as $\mathcal{O}(B\times C_\mathrm{out}\times H\times W)$,
the average pooling layer can be represented as
$\mathcal{O}(C_\mathrm{in}\times K_\mathrm{avg}^2\times H\times W)$,
and of the max pooling layer can be represented as
$\mathcal{O}(C_\mathrm{in}\times K_\mathrm{max}^2\times H\times W)$.
Hence, the time complexity of the texture feature extraction can be denoted as
\begin{equation}
	\begin{aligned}
		& \mathcal{O}\left(\sum_{u=1}^UC_\mathrm{in}^u\times C_\mathrm{out}^u\times (K_\mathrm{conv}^{u})^{2}\times H^u\times W^u\right)\\
		& + \mathcal{O}\left(\sum_{u=1}^UB\times C_\mathrm{out}^u\times H^u\times W^u\right) \\
		& + \mathcal{O}(C_\mathrm{in} \times K_\mathrm{avg}^2 \times H \times W) + \mathcal{O}(C_\mathrm{in} \times K_\mathrm{max}^2 \times H \times W).
	\end{aligned}
\end{equation}

The time complexity of position features mainly comes from position encoding and the TE. Therein the time complexity of position encoding is $\mathcal{O}(\ell\times d)$, where $\ell$ is the length of the sequence, while $d$ is the length of the word vector.
The time complexity of TE mainly comes from the matrix operation among the linear transformation and the query ($\boldsymbol{Q}$), the key ($\boldsymbol{K}$), and the value ($\boldsymbol{V}$). The time complexity of linear transformation can be expressed as $\mathcal{O}(\ell\times d^{2})$, whereas the time complexity of matrix operations can be denoted as $\mathcal{O}(\ell^{2}\times d)$. Therefore, the overall time
complexity can be denoted as
\begin{equation}
	\mathcal{O}(\ell\times d^{2} + \ell^{2}\times d).
\end{equation}
Hence, the time complexity of the entire feature extraction part can be represented as
\begin{equation}
\mathcal{O}(\ell\times d)+\mathcal{O}(\ell\times d^{2} + \ell^{2}\times d).
\end{equation}
\begin{table*}[!ht]
	\centering
	\caption{Model parameters, computation, and inference latency}
	\begin{tabular}{ccccc}
		\toprule
		\bfseries Model &
		\bfseries Total params &
		\bfseries Params Size (MB) &
		\bfseries FLOPs &
		\bfseries Inference Time (ms) \\
		\midrule
		Open-RFNet & 205812624 & 785 & 1034748864 & 54.21 \\
		ResNet & 700816 & 2.67 & 471785920 & 14.41 \\
		\bottomrule
	\end{tabular}
	\label{table_para}
\end{table*}

With the quantitative analysis, we can calculate the parameter count and the computational complexity of the Open-RFNet, as shown in Table~\ref{table_para}. It can be found that by incorporating the position feature extraction module based on ResNet, the total number of parameters and their size increased by nearly 292 times, whereas the total number of floating-point operations (FLOPs) required for a single forward propagation of the model increased by 1.19 times, i.e., the final FLOPs are 2.19 times the original.

As UAV RF recognition is a real-time application, we further analyze the end-to-end response time of the system, including: (i) signal receiving time, (ii) signal preprocessing time with slicing, denoising, and STFT operations, and (iii) model inference time. In practice, the receiving time is determined by the sampling window. In the DroneRFa dataset used in this study, each sample contains approximately 300,000 IQ samples, collected at a sampling rate of 100 MS/s, corresponding to a signal receiving time of about 3 ms. The preprocessing stage is dominated by the STFT operation, with an average latency of 21.76 ms per sample. For model inference, the inference latency is 14.41 ms for ResNet and 54.21 ms for Open-RFNet. Although the latter incurs slightly higher latency, both remain within the millisecond range. Overall, the entire processing pipeline operates at the millisecond level, which can meet the real-time requirements of UAV monitoring and identification applications.
\section{Experiments}
\label{sec:7}
\subsection{Experimental Setting}

To compare the performances with the state-of-the-art methods \cite{bendale2016towards,ge2017generative,yu2024open,wang2023uncertainty}, we use the public dataset DroneRFa as the benchmark dataset, which includes the RF signals of 24 common types of UAVs and flight controllers, and one type of background signal as a reference. The specific settings are shown in Table~\ref{table_1}. The RF signal collection covers three Industrial Scientific Medical (ISM) bands, namely 915 MHz, 2.4 GHz and 5.8 GHz, which are collected by a universal software radio peripheral (USRP) and monitored simultaneously in a dual-channel mode. Therein, 15 types of UAV signals are collected indoors in hovering mode and 5 meters away from the receiver. In addition, the signals of 9 types of UAVs are collected outdoors and are divided into three classifications according to the actual flight distance: near, medium, and far. However, some signals at medium and long distances are very weak or even non-existent, which makes these data samples untrainable. Therefore, considering the availability of the data, we select the signals collected at near distances and indoors for experiment evaluation. Finally, we slice the I/Q data into small segments of 3 ms with 300,000 data points and obtain a time-frequency matrix of (775, 775) via the STFT.

\begin{table*}[!ht]
	\centering
	\caption{The classes of the used dataset}
	\begin{tabular}{>{\centering\arraybackslash}p{1.5cm}>{\centering\arraybackslash}p{1.5cm}>{\centering\arraybackslash}p{3cm}>{\centering\arraybackslash}p{4cm}}
		\toprule
		\bfseries ID & \bfseries Label & \bfseries Class & \bfseries Spectrum Band \\
		\midrule
		A & T0000 & Backgroundd & Both 2.4G and 5.8G \\
		B & T0001 & DJI Phantom 3 & Only 2.4G \\
		C & T0010 & DJI Phantom 4 Pro & Both 2.4G and 5.8G \\
		D & T0011 & DJI MATRICE 200 & Both 2.4G and 5.8G \\
		E & T0100 & DJI MATRICE 100 & Only 2.4G \\
		F & T0101 & DJI Air 2S & Both 2.4G and 5.8G \\
		G & T0110 & DJI Mini 3 Pro & Both 2.4G and 5.8G \\
		H & T0111 & DJI Inspire 2 & Both 2.4G and 5.8G \\
		I & T1000 & DJI Mavic Pro & Only 2.4G \\
		J & T1001 & DJI Mini 2 & Both 2.4G and 5.8G \\
		K & T1010 & DJI Mavic 3 & Only 5.8G \\
		L & T1011 & DJI MATRICE 300 & Both 2.4G and 5.8G \\
		M & T1100 & DJI Phantom 4 Pro RTK & Only 5.8G \\
		N & T1101 & DJI MATRICE 30T & Both 2.4G and 5.8G \\
		O & T1110 & DJI AVATA & Both 2.4G and 5.8G \\
		P & T1111 & DJI DIY & Only 5.8G \\
		Q & T10000 & DJI MATRICE 600 Pro & Only 2.4G \\
		R & T10001 & VBar & Only 2.4G \\
		S & T10010 & FrSky X20 & Both 2.4G and 915M \\
		T & T10011 & Futaba T6IZ & Only 2.4G \\
		U & T10100 & Taranis Plus & Only 915M \\
		V & T10101 & RadioLink AT9S & Only 2.4G \\
		W & T10110 & Futaba T14SG & Only 2.4G \\
		X & T10111 & Skydroid T12 & Only 2.4G \\
		Y & T11000 & Skydroid T10 & Only 2.4G \\
		\bottomrule
		\label{table_1}
	\end{tabular}
\end{table*}
\begin{figure}[!h]
	\centering
	\includegraphics[width=3.1in]{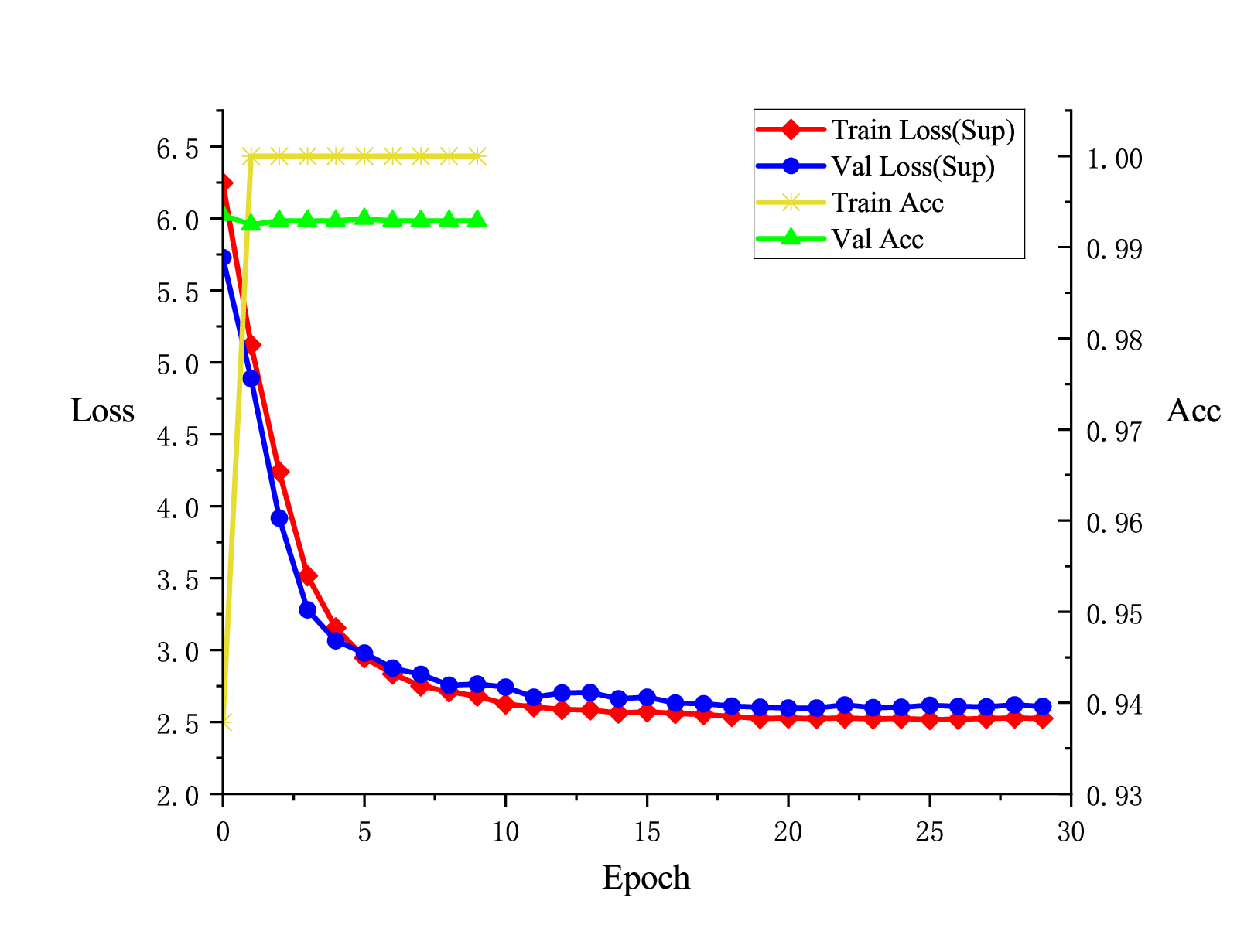}
	\caption{Training loss with respect to the epoch.}
	\label{fig_Loss}
\end{figure}
The division of the known and the unknown classifications is shown in Table~\ref{table_2}. A total of 20 known classifications and 5 unknown classifications are selected. The model is trained using a Tesla V100 SXM2 32GB graphics card. The sampling rate is 100 MS/s, the batchsize is 128 and the optimizer is Adam. The maximum number of epochs for contrastive learning pre-training is set to be 30, while the subsequent classification fine-tuning training lasts for 10 epochs. The learning rate is scheduled by using the cosine annealing algorithm. Figure~\ref{fig_Loss} shows that the proposed Open-RFNet can converge normally in both stages and has no overfitting.

\begin{table}[!ht]
	\centering
	\caption{The division of known classes and unknown classes in the dataset}
	\begin{tabular}{c|c}
		\hline
		\multicolumn{2}{c}{Splitting setting (A $\sim$ X refer to class indexes in Table \ref{table_1})} \\ \hline
		known classes (20) & A C D E F G H I J K L M N O P Q S U V X  \\ \hline
		unknown classes (5) & B R T W Y  \\ \hline
	\end{tabular}
	\label{table_2}
\end{table}

\subsection{Evaluation Indicator}

We use the known accuracy rate (KAR), unknown accuracy rate (UAR), known precision (KP), unknown precision (UP), and performance gap (GAP) to comprehensively evaluate the proposed method, which are given by
\begin{equation}
	\begin{aligned}
		\mathrm{KAR}&=\frac{CK}{TK + FU}, \\
		\mathrm{UAR}&=\frac{TU}{TU + FK}, \\
		\mathrm{KP}&=\frac{CK}{TK + FK}, \\
		\mathrm{UP}&=\frac{TU}{TU + FU}, \\
		\mathrm{GAP}&=|\mathrm{UAR}-\mathrm{KAR}|,
	\end{aligned}
\end{equation}
where $TK$, $TU$, $FK$ and $FU$ are the numbers of test samples that are identified as true known, true unknown, false known, and false unknown, respectively. The number of correctly known samples, i.e., $CK$, represents the total number of the known test samples that are correctly classified into the exact known classifications. The GAP is defined as the absolute value of the difference between UAR and KAR, which is used to measure the balance between the recognition performance of the known classification and the unknown classification.

\subsection{Ablation Experiment}
Before presenting the ablation results, we clarify the definitions of the comparison models used in this section. SupResNet refers to a ResNet backbone equipped only with the supervised contrastive learning module. TransNet denotes a model where the ResNet is replaced by the proposed TE, without using supervised contrastive learning. These two models are designed to isolate the effects of the TE and the supervised contrastive learning, respectively, and serve as baseline variants for evaluating the contribution of each component in the proposed Open-RFNet. To verify the effectiveness of the constructed TE, supervised contrastive learning and the denoising preprocessing module in the proposed Open-RFNet, the ablation experiments are conducted for open-set recognition. We first verify the importance of the introduced TE and supervised contrastive learning, which as illustrated in Table~\ref{table_5}. The red values in the parentheses of the table represent the differences of the component missing models compared to Open-RFNet. It can be seen that when the TE or the supervised contrastive learning is introduced alone on the basis of ResNet, no performance improvement can be achieved. However, if both the Transformer structure and the supervised contrastive learning are added simultaneously, a significant improvement in recognition performance can be obtained. Specifically, the KAR of TransNet with the TE introduced alone is only 95.79$\%_{}$, and the UAR is 87.6$\%_{}$. Compared with the ResNet, they decrease by 0.73$\%_{}$ and 3.68$\%_{}$ respectively. This indicates that the recognition performance deteriorates both in the known classifications and the unknown classifications, especially for the latter. After introducing the supervised contrastive learning only, the KAR and UAR also decrease by 0.77$\%_{}$ and 0.8$\%_{}$ respectively. Yet,  after adding both the TE and the supervised contrastive learning, the TKR decreases by 1.4$\%_{}$, while the TUR increases by 4.8$\%_{}$. This implies that we sacrifice a certain recognition performance of the known classifications in exchange for a huge performance improvement in the recognition of the unknown classifications. From the perspective of the overall recognition performance, we find that the proposed Open-RFNet is more balanced and has better performance in the recognition of both the known and the unknown classifications. Additionally,  we verify the effectiveness of the proposed denoising preprocessing. The results are shown in Table~\ref{table_6}. The red values in the parentheses of the table represent the differences between the denoised model and the noisy model. It can be clearly observed that after data denoising, the open-set recognition performance has been significantly improved compared with the noisy data. Specifically, in the noisy data, the IG-OpenMax algorithm performs poorly on each model, with the highest accuracy for the unknown classification recognition being only 66.08$\%_{}$ and the lowest being 52.16$\%_{}$, indicating a relatively low recognition performance. After denoising preprocessing, the recognition performance for the unknown classifications increased significantly, where the highest is 96.08$\%_{}$ and the lowest is 87.6$\%_{}$.

\begin{table*}[!ht]
	\centering
	\caption{Comparison of Open-RFNet with SupResNet, TransNet, and ResNet}
	\begin{tabular}{c|c|c|c|c|c}
		\hline
		\multicolumn{2}{c|}{\diagbox{Indicator}{Models~~~~~~~~~}} & \textbf{Open-RFNet} & SupResNet & TransNet & ResNet \\ \hline
		\multirow{20}{*}{Known accuracy}
		& T0000(A) & 0.9008 & 0.9463 & 0.9587 & 0.9752 \\ \cline{2-6}
		& T0010(C) & 0.9835 & 0.9628 & 0.9835 & 0.9752 \\ \cline{2-6}
		& T0011(D) & 0.9504 & 0.9628 & 0.9669 & 0.9669 \\ \cline{2-6}
		& T0100(E) & 0.938 & 0.9711 & 0.9711 & 0.9793 \\ \cline{2-6}
		& T0101(F) & 0.9876 & 1 & 0.9917 & 0.9876 \\ \cline{2-6}
		& T0110(G) & 0.9751 & 0.9751 & 0.9793 & 0.9793 \\ \cline{2-6}
		& T0111(H) & 0.9298 & 0.9504 & 0.9463 & 0.9628 \\ \cline{2-6}
		& T1000(I) & 0.9834 & 0.9917 & 0.9793 & 0.9834 \\ \cline{2-6}
		& T1001(J) & 0.9793 & 0.9793 & 0.9834 & 0.9793 \\ \cline{2-6}
		& T1010(K) & 0.9502 & 0.9544 & 0.9461 & 0.9668 \\ \cline{2-6}
		& T1011(L) & 0.9834 & 0.9793 & 0.9751 & 0.9834 \\ \cline{2-6}
		& T1100(M) & 0.9959 & 1 & 0.9917 & 1 \\ \cline{2-6}
		& T1101(N) & 0.971 & 0.9585 & 0.9502 & 0.9751 \\ \cline{2-6}
		& T1110(O) & 0.9032 & 0.9032 & 0.9078 & 0.9078 \\ \cline{2-6}
		& T1111(P) & 0.9087 & 0.917 & 0.9502 & 0.9917 \\ \cline{2-6}
		& T10000(Q) & 0.9793 & 0.9917 & 0.971 & 0.9876 \\ \cline{2-6}
		& T10010(S) & 0.905 & 0.9008 & 0.9091 & 0.9008 \\ \cline{2-6}
		& T10100(U) & 0.9959 & 0.9959 & 0.9835 & 0.9917 \\ \cline{2-6}
		& T10101(V) & 0.9008 & 0.9008 & 0.905 & 0.9008 \\ \cline{2-6}
		& T10111(X) & 0.9004 & 0.9046 & 0.9046 & 0.9046 \\ \hline
		\multirow{5}{*}{Unknown accuracy}
		& T0001(B) & 0.976 & 0.94 & 0.856 & 0.952 \\ \cline{2-6}
		& T10001(R) & 0.9 & 0.868 & 0.792 & 0.9 \\ \cline{2-6}
		& T10011(T) & 0.972 & 0.92 & 0.948 & 0.892 \\ \cline{2-6}
		& T10110(W) & 0.968 & 0.984 & 0.884 & 0.948 \\ \cline{2-6}
		& T10000(Y) & 0.988 & 0.812 & 0.9 & 0.872 \\ \hline
		\multicolumn{2}{c|}{Closed Acc} & 0.994 & 0.9929 & 0.9929 & 0.9958 \\ \hline
		\multicolumn{2}{c|}{KAR} & 0.9512 & 0.9575 \textcolor{red}{(+0.0063)} & 0.9579 \textcolor{red}{(+0.0067)} & 0.9652 \textcolor{red}{(+0.014)} \\ \hline
		\multicolumn{2}{c|}{UAR} & 0.9608 & 0.9048 \textcolor{red}{(-0.056)} & 0.876 \textcolor{red}{(-0.0848)} & 0.9128 \textcolor{red}{(-0.048)} \\ \hline
		\multicolumn{2}{c|}{KP} & 0.9887 & 0.9735 & 0.9657 & 0.9747 \\ \hline
		\multicolumn{2}{c|}{UP} & 0.8386 & 0.851 & 0.8494 & 0.8797 \\ \hline
		\multicolumn{2}{c|}{GAP} & \textbf{0.0096} & \textbf{0.0527} & \textbf{0.0819} & \textbf{0.0524} \\ \hline
	\end{tabular}
	\label{table_5}
\end{table*}

\begin{table*}[!ht]
	\centering
	\caption{Comparison between the denoised and the noisy data}
	\begin{tabular}{c|c|c|c|c|c|c}
		\hline
		\multicolumn{2}{c|}{\diagbox{Indicator}{Algorithms}} & KAR & UAR & KP & UP & GAP \\ \hline
		\multirow{6}{*}{Noisy}
		& Open-RFNet & 0.946 & 0.6504 & 0.9087 & 0.772 & 0.2956 \\ \cline{2-7}
		& SupResNet & 0.9525 & 0.6608 & 0.9122 & 0.7957 & 0.2917 \\ \cline{2-7}
		& TransNet & 0.9612 & 0.5216 & 0.8803 & 0.8059 & 0.4396 \\ \cline{2-7}
		& ResNet & 0.9641 & 0.5632 & 0.8905 & 0.8253 & 0.4009 \\ \cline{2-7}
 \hline
		\multirow{6}{*}{Denoised}
		& Open-RFNet &  0.9512 (\textcolor{red}{+0.0052}) & 0.9608 (\textcolor{red}{+0.3104}) &  0.9887 & 0.8365 & 0.0096 \\ \cline{2-7}
		& SupResNet &  0.9575 (\textcolor{red}{+0.0050}) & 0.9048 (\textcolor{red}{+0.2440}) &  0.9735 & 0.851 & 0.0527 \\ \cline{2-7}
		&  TransNet & 0.9579 (\textcolor{red}{-0.0033}) &  0.876 (\textcolor{red}{+0.3544}) &  0.9657 & 0.8494 & 0.0819 \\ \cline{2-7}
		&  ResNet &  0.9652 (\textcolor{red}{+0.0011}) & 0.9128 (\textcolor{red}{+0.3496}) &  0.9747 & 0.8797 & 0.0524 \\ \cline{2-7}
		\hline
	\end{tabular}
	\label{table_6}
\end{table*}

\subsection{Comparison of Different Algorithms}
Here, we focus on comparing different state-of-the-art open-set recognition algorithms.
To ensure clarity, we first describe the models involved in this comparison.
The Open-RFNet is the proposed model based on the IG-OpenMax algorithm.
Open-RFNet-B follows the same feature extraction pipeline but replaces IG-OpenMax with the classical OpenMax algorithm \cite{bendale2016towards}.
Open-RFNet-G further replaces the open-set module with the G-OpenMax method \cite{ge2017generative}.
In addition, S3R \cite{yu2024open} and UIOS \cite{wang2023uncertainty} are two state-of-the-art open-set recognition methods, and we reproduce their models strictly following the configurations described in their original papers.
Furthermore, Open-RFNet-S3R-O and Open-RFNet-UIOS-O denote the variants where the open-set module in Open-RFNet is replaced by the S3R and UIOS algorithms, respectively, while keeping the feature extraction network unchanged.
Based on these definitions, we compare the proposed Open-RFNet with Open-RFNet-B, Open-RFNet-G, the S3R model, and the UIOS model to evaluate open-set recognition performance under consistent settings. First, we compare the proposed Open-RFNet with the open-set recognition S3R and UIOS models, and the results are shown in Table~\ref{table_7}. The red values in the parentheses of the table represent the performance differences. It can be found that the KUR of the S3R model is extremely high, and can almost recognize the unknown classes with 100$\%_{}$ accuracy, but it performs poorly in terms of KAR, which is only 90.86$\%_{}$. Moreover, there is a large difference in the recognition accuracy of various classes. In particular, the recognition accuracies of the classes T0011, T0100, and T0111 are all less than 80$\%_{}$. On the contrary, Open-RFNet performs well on the known classes, with each recognition accuracy of class above 90$\%_{}$ and an overall recognition accuracy of 95.12$\%_{}$. For the UIOS model, its KAR is close to that of the the Open-RFNet, but the UAR is 2.16$\%_{}$ lower than that of Open-RFNet, which indicates that its recognition ability for the unknown samples is weaker than that of the Open-RFNet. Next, to further evaluate the performance of the proposed IG-Openmax algorithm, we compare the Open-RFNet with Open-RFNet-B, Open-RFNet-G, Open-RFNet-S3R-O, and Open-RFNet-UIOS-O algorithms, where the abbreviations S3R-O and UIOS-O denote the open-set algorithms of the S3R and the UIOS models. The results are shown in Table~\ref{table_8}. The red values in the parentheses of the table represent the performance differences. For the Open-RFNet, compared with the OpenMax algorithm, the IG-OpenMax algorithm increases the UAR by 2.24$\%_{}$, and compared with the G-OpenMax algorithm, the UAR increases by 2.96$\%_{}$, with the KAR also remaining almost unchanged. The IG-OpenMax algorithm can effectively improve the recognition performance of the unknown classes without compromising the recognition performance of the known classes. Compared with the S3R-O and UIOS-O algorithms, the IG-OpenMax algorithm shows improvements in both the KAR and UAR. Moreover, the UIOS-O algorithm cannot recognize the unknown classes based on the Open-RFNet structure, where the performance gap reaches 93.98\%. The significant performance differences between the UIOS-O and the S3R-O algorithms arise from both the model structure and the loss function, as well as the training strategy. The UIOS-O and S3R-O algorithms have to revise the loss functions for subsequent open-set recognition during model training, whereas the proposed IG-OpenMax algorithm only uses the commonly cross-entropy loss function, which is more versatile and has lower training overhead.
\subsection{Explanation of IG-OpenMax}

\begin{figure}[!t]
	\centering
	\subfloat[IG-OpenMax]{\includegraphics[width=0.245\textwidth, height=0.245\textwidth]{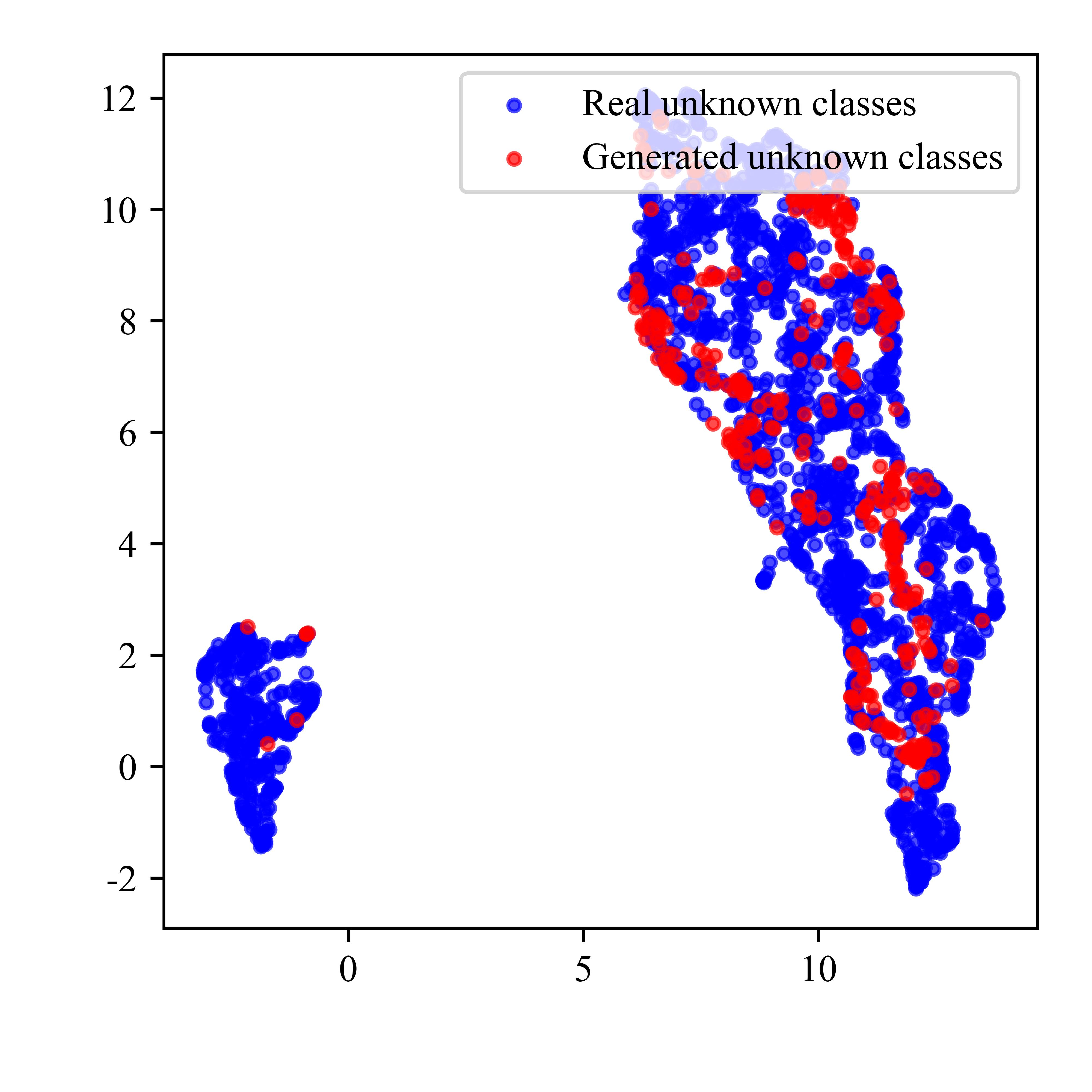} \label{fig_5_1}}
	\subfloat[G-OpenMax]{\includegraphics[width=0.245\textwidth, height=0.245\textwidth]{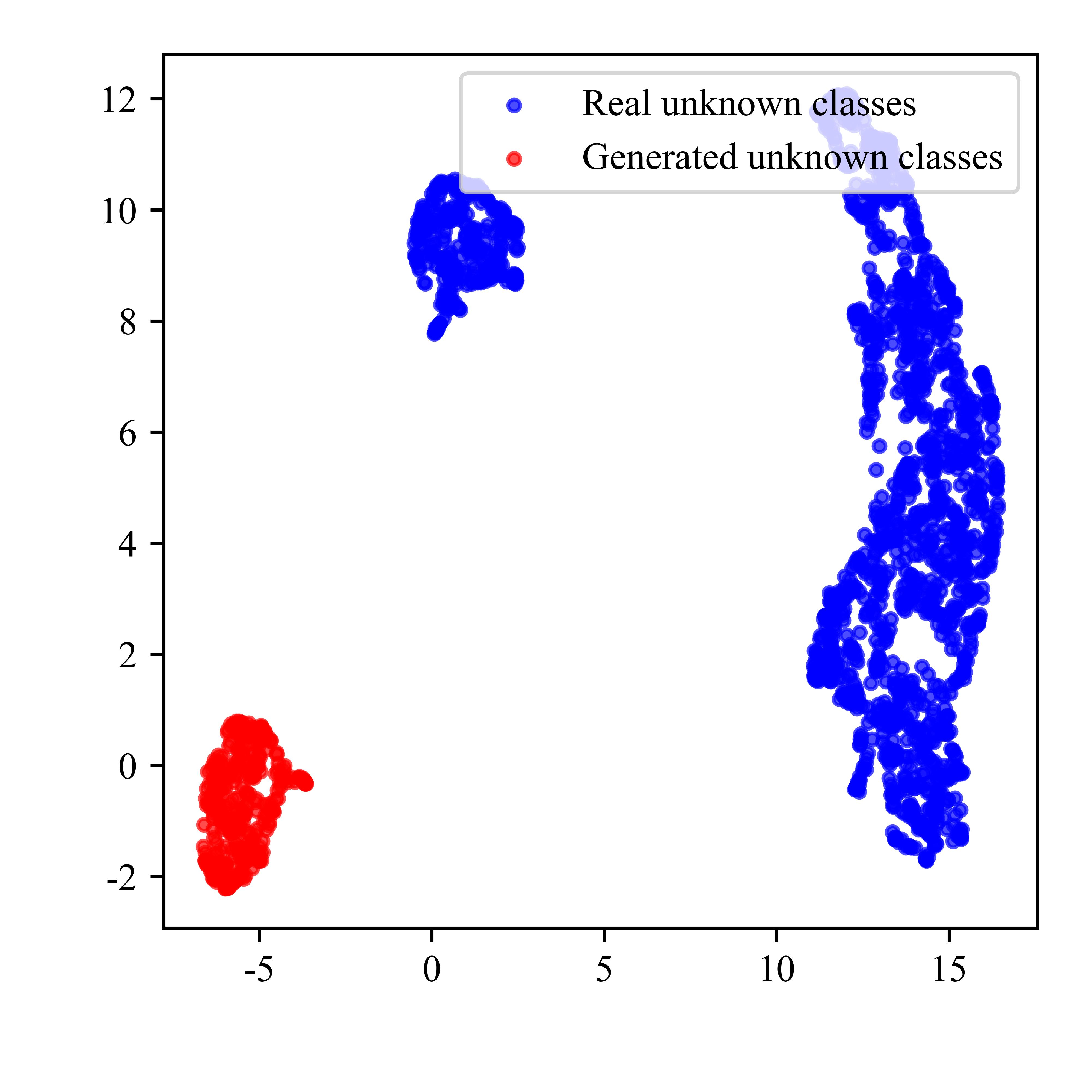} \label{fig_5_2}}
	\caption{Visualization of the distribution of known classifications.}
	\label{fig_5}
\end{figure}

Figure~\ref{fig_5} provides a feature-level visualization of the real unknown class samples and the generated unknown samples under the G-OpenMax and IG-OpenMax algorithms.
In the IG-OpenMax case, i.e., Fig.~\ref{fig_5_1}, the generated unknown samples almost fully overlap with the real unknown samples in the feature space. This confirms that, by freezing the feature extractor and retraining only the classification head, the original feature structure learned by Open-RFNet-C is preserved.
Within this unchanged feature space, the misclassified generated samples tend to lie in boundary regions, where real unknown samples are also likely to appear. As a result, the generated samples can faithfully approximate the true unknown class distribution, enabling the classifier to learn more effectively decision boundaries for open-set recognition.

In contrast, the G-OpenMax visualization in Fig.~\ref{fig_5_2} shows that the generated unknown samples form a cluster that is entirely separated from the real unknown samples. This separation arises because G-OpenMax retrains the entire network, which causes the feature space to shift significantly. In the new feature space, the model is able to easily distinguish generated samples from real data, which prevents the generated unknown samples from serving as valid approximations of the real unknown classes.

These observations demonstrate that freezing the feature extractor is crucial. Therein, the IG-OpenMax maintains the discriminative structure of the original feature space, ensures similarity between generated and real unknown samples at the feature level, and ultimately leads to substantially improved open-set recognition performance.

\begin{table*}[!ht]
	\centering
	\caption{Comparison of the proposed Open-RFNet with S3R model and UIOS model}
	\begin{tabular}{c|c|c|c|c}
		\hline
		\multicolumn{2}{c|}{\diagbox{Indicator}{Models~~~~~~~}} & \textbf{Open-RFNet} & S3R \cite{yu2024open}& UIOS \cite{wang2023uncertainty}\\ \hline
		\multirow{20}{*}{Known accuracy}
		& T0000(A) & 0.9008 & 0.8843 & 0.8843 \\ \cline{2-5}
		& T0010(C) & 0.9835 & 0.8471 & 0.9174 \\ \cline{2-5}
		& T0011(D) & 0.9504 & 0.7603 & 0.9504 \\ \cline{2-5}
		& T0100(E) & 0.938 & 0.7355 & 0.9008 \\ \cline{2-5}
		& T0101(F) & 0.9876 & 0.9959 & 1 \\ \cline{2-5}
		& T0110(G) & 0.9751 & 0.9502 & 0.9585 \\ \cline{2-5}
		& T0111(H) & 0.9298 & 0.7438 & 0.9132 \\ \cline{2-5}
		& T1000(I) & 0.9834 & 0.9917 & 1 \\ \cline{2-5}
		& T1001(J) & 0.9793 & 0.971 & 0.9917 \\ \cline{2-5}
		& T1010(K) & 0.9502 & 0.9212 & 0.9378 \\ \cline{2-5}
		& T1011(L) & 0.9834 & 0.9627 & 0.9917 \\ \cline{2-5}
		& T1100(M) & 0.9959 & 0.9585 & 0.9834 \\ \cline{2-5}
		& T1101(N) & 0.971 & 0.9668 & 0.9834 \\ \cline{2-5}
		& T1110(O) & 0.9032 & 0.8986 & 0.8986 \\ \cline{2-5}
		& T1111(P) & 0.9087 & 0.8921 & 0.8548 \\ \cline{2-5}
		& T10000(Q) & 0.9793 & 0.9419 & 0.9917 \\ \cline{2-5}
		& T10010(S) & 0.905 & 0.9256 & 0.9463 \\ \cline{2-5}
		& T10100(U) & 0.9959 & 0.9917 & 1 \\ \cline{2-5}
		& T10101(V) & 0.9008 & 0.905 & 0.9587 \\ \cline{2-5}
		& T10111(X) & 0.9004 & 0.9295 & 0.9502 \\ \hline
		\multirow{5}{*}{Unknown accuracy}
		& T0001(B) & 0.976 & 1 & 0.948 \\ \cline{2-5}
		& T10001(R) & 0.9 & 0.996 & 0.94 \\ \cline{2-5}
		& T10011(T) & 0.972 & 0.976 & 0.944 \\ \cline{2-5}
		& T10110(W) & 0.968 & 0.996 & 0.984 \\ \cline{2-5}
		& T10000(Y) & 0.988 & 0.992 & 0.88 \\ \hline
		\multicolumn{2}{c|}{Close Acc} & 0.994 & 0.9942 & 0.9963 \\ \hline
		\multicolumn{2}{c|}{KAR} & 0.9512 & 0.9086 \textcolor{red}{(-0.0426)} & 0.9512 \textcolor{red}{(0)} \\ \hline
		\multicolumn{2}{c|}{UAR} & 0.9608 & 0.992 \textcolor{red}{(+0.0312)} & 0.9392 \textcolor{red}{(-0.0216)} \\ \hline
		\multicolumn{2}{c|}{KP} & 0.9887 & 0.9977 & 0.9836 \\ \hline
		\multicolumn{2}{c|}{UP} & 0.8386 & 0.7385 & 0.8338 \\ \hline
		\multicolumn{2}{c|}{GAP} & \textbf{0.0096} & \textbf{0.0834} & \textbf{0.012} \\ \hline
	\end{tabular}
	\label{table_7}
\end{table*}

\begin{table*}[!ht]
	\centering
	\caption{Comparison among open-set recognition methods}
	\begin{tabular}{c|c|c|c|c|c}
		\hline
		\diagbox{Indicator}{Methods~~~~} & KAR & UAR & KP & UP & GAP \\ \hline
		\textbf{Open-RFNet}  &  0.9512 &  0.9608 &  0.9887 &  0.8386 & \textbf{0.0096} \\ \cline{1-6}
	Open-RFNet-B	 & 0.9523 (\textcolor{red}{+0.0011}) & 0.9384 (\textcolor{red}{-0.0224}) & 0.9826 & 0.839 & \textbf{0.0139} \\ \cline{1-6}
	Open-RFNet-G & 0.9529 (\textcolor{red}{+0.0017}) & 0.9312 (\textcolor{red}{-0.0296}) & 0.9811 & 0.8386 & \textbf{0.0217} \\ \cline{1-6}
	Open-RFNet-S3R-O	 & 0.9462 (\textcolor{red}{-0.0050}) & 0.8864 (\textcolor{red}{-0.0744}) & 0.969 & 0.8129 & \textbf{0.0598} \\ \cline{1-6}
	Open-RFNet-UIOS-O & 0.9398 (\textcolor{red}{-0.0114}) & 0 (\textcolor{red}{-0.9608}) & 0.7792 & 0 & \textbf{0.9398} \\ \hline
	\end{tabular}
	\label{table_8}
\end{table*}

\begin{figure*}[!t]
	\centering
	\subfloat[ResNet]{\includegraphics[width=0.23\textwidth, height=0.23\textwidth]{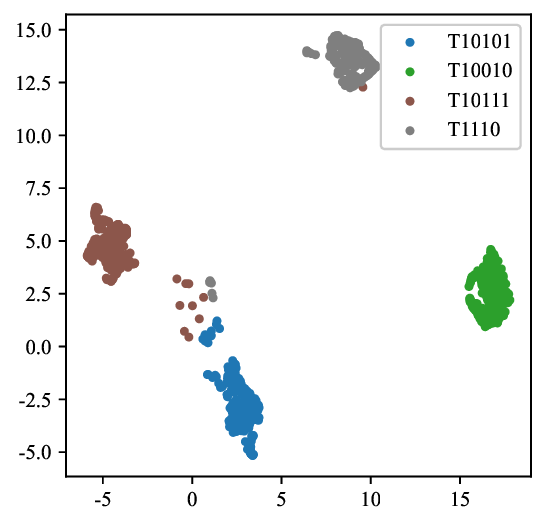} \label{fig6_1}}
	\subfloat[SupResNet]{\includegraphics[width=0.23\textwidth, height=0.23\textwidth]{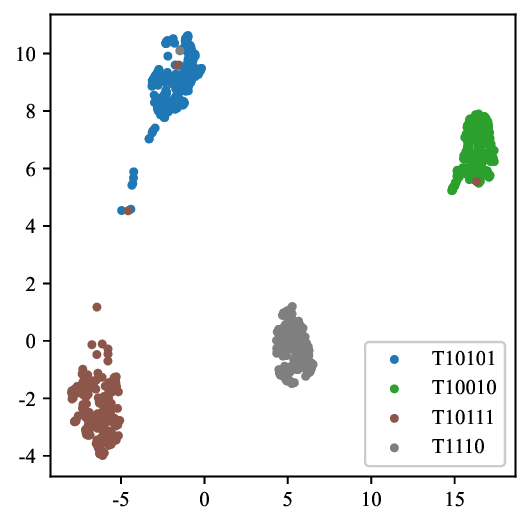} \label{fig6_2}}
	\subfloat[TransNet]{\includegraphics[width=0.23\textwidth, height=0.23\textwidth]{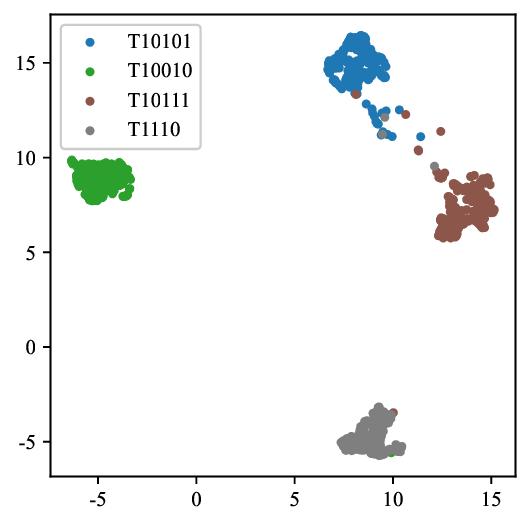} \label{fig6_3}}
	\subfloat[Open-RFNet]{\includegraphics[width=0.23\textwidth, height=0.23\textwidth]{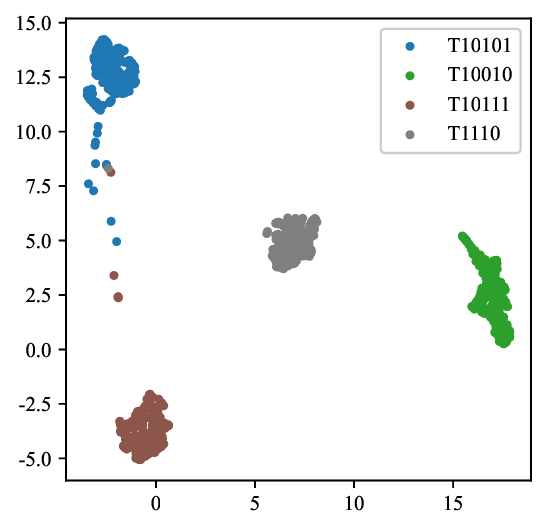} \label{fig6_4}}
	\caption{Visualization of the distribution of known classifications.}
	\label{fig_6}
\end{figure*}

\subsection{Explanation of Proposed Model with Visualization}

\begin{figure*}[!t]
	\centering
	\subfloat[ResNet]{\includegraphics[width=0.23\textwidth, height=0.23\textwidth]{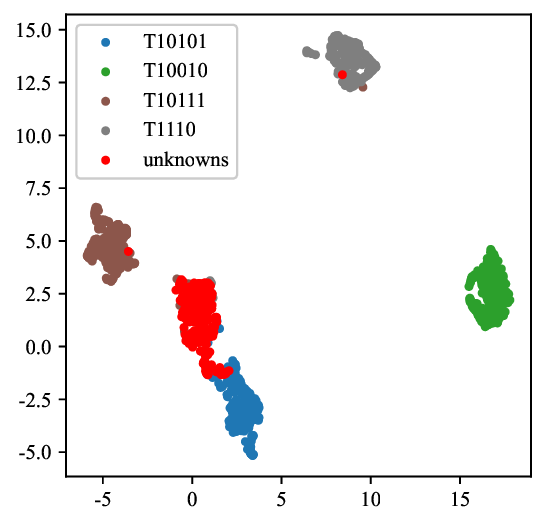} \label{fig7_1}}
	\subfloat[SupResNet]{\includegraphics[width=0.23\textwidth, height=0.23\textwidth]{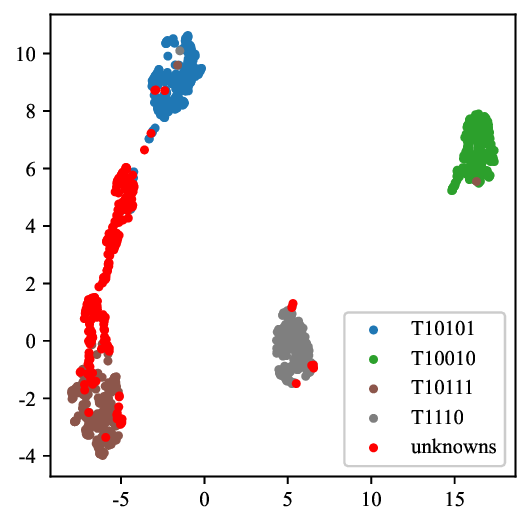} \label{fig7_2}}
	\subfloat[TransNet]{\includegraphics[width=0.23\textwidth, height=0.23\textwidth]{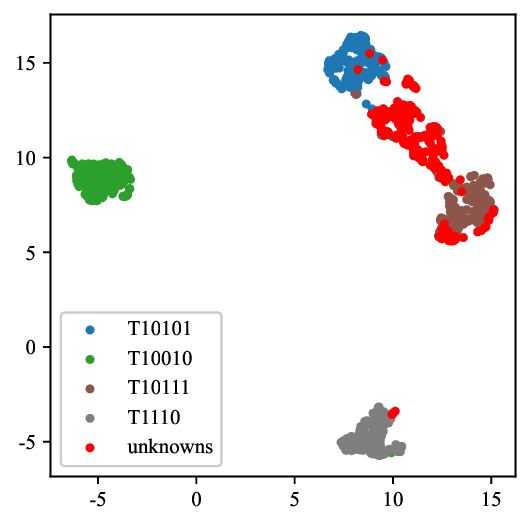} \label{fig7_3}}
	\subfloat[Open-RFNet]{\includegraphics[width=0.23\textwidth, height=0.23\textwidth]{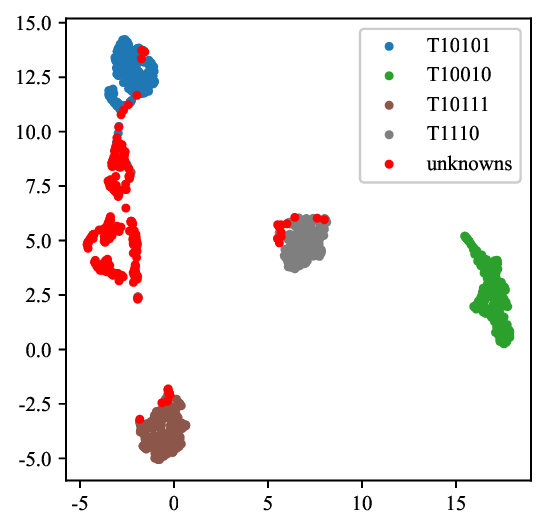} \label{fig7_4}}
	\caption{Visualization of the distributions of known and unknown classifications.}
	\label{fig_7}
\end{figure*}
The performance among the models can be explained by visualizing the
AVs. In Fig.~\ref{fig_6} and Fig.~\ref{fig_7}, we illustrate a visualization of the AVs for the four models, which are Open-RFNet, SupResNet, TransNet, and ResNet. To clearly demonstrate the performance differences, the models are completely generated from the known classification, and no generated unknown classification samples are added. This is because the generated unknown classification samples selected by different models are inconsistent, which can avoid the influence caused by inconsistent training data. For the unknown classification, only T10110 is selected, and for the known classifications, T10101, T10010, T10111, and T1110 are selected.

First, we focus on ResNet and SupResNet. It can be clearly observed that the distance between classes has significantly increased, whether in the separation of known classes or the distribution of unknown classes. This reflects the characteristic of supervised contrastive learning in pulling apart heterogeneous samples. However, the overlap between the unknown classes and the known classes still exists, indicating that the known and the unknown classes cannot be distinguished relying solely on texture features. Then, we focus on ResNet and TransNet, both of which are trained using the cross-entropy loss function. It can be found that the two models are highly similar in the distribution of known categories. In the joint distribution of the known and the unknown classes, and although there are slight differences between them, the overlap between unknown and known classes is similar. This indicates that when training with the cross-entropy loss function, an imbalance in multi-feature optimization occurs; that is, the model pays much attention to the texture features while ignoring the time-frequency position features, leading to the similarity in the visualization results. Then, for the ResNet and the Open-RFNet, it can be seen that after adding both time-frequency position features and supervised contrastive learning, the known classes and the unknown classes are clearly distinguished. Although there are still a few unknown class samples close to the known classes, they are all at the edge of the distribution, which can be easily recognized by open-set recognition algorithms. This shows that the supervised contrastive learning effectively solves the problem of imbalance in multi-feature optimization, enabling the model to fully learn the position features, thereby achieving enhanced UAV RF recognition performance.
\section{Conclusion}
\label{sec:8}
This paper proposed an MD-SupContrast framework for UAV recognition for an open-set scenario. The proposed Open-RFNet utilized the ResNet and the TE modules to extract texture features and time-frequency position features of the RF signals, realized feature fusion, and incorporated supervised contrastive learning to further optimize the features. In addition, an IG-OpenMax algorithm was also conceived to balance the recognition performance between the closed-set and the open-set. When introducing the generated unknown samples, the feature extraction part is frozen, and only the classification head is retrained, enabling the model to better distinguish between the known and the unknown classifications. Compared with the advanced methods, the superiority of the proposed method was demonstrated. Specifically, we can recognize 25 UAV types with recognition accuracy 95.12\% in closed-set and 96.08\% in open-set, respectively, while the
performance gap
is less than 1\%. We found that the complementary recognition mechanism between
the time-frequency
domain position features and the supervised contrastive learning
effectively improves the distribution relationship between the
samples of the known and the unknown classifications, thereby significantly enhancing the UAV RF open-set recognition performance.

It is worth noting that the proposed framework is developed under the single UAV emitter assumption, where only one UAV RF emission is present within the given time. However, simultaneously recognizing multiple UAV emitters is a meaningful and challenging problem, as superimposed RF signals will introduce severe mutual interference and significantly complicate the extraction of discriminative features. Thus, in our future research, we will focus on multiple UAV RF recognition and further pursue their open-set recognition.

\bibliographystyle{IEEEtran}
\bibliography{paper}

\newpage

\vfill

\end{document}